\documentclass[final,5p,times,twocolumn]{elsarticle}
\usepackage{multirow}
\usepackage{helvet}
\usepackage{courier}
\usepackage{booktabs}
\usepackage{makecell}
\usepackage{threeparttable}
\usepackage{bm}
\usepackage{cite}
\usepackage{comment}
\usepackage{float}
\usepackage{amsmath}
\usepackage{amssymb,graphicx,ntheorem}
\usepackage{epstopdf}
\usepackage{color,url}
\usepackage{algorithm}
\usepackage{algpseudocode}
\usepackage{array}
\usepackage{epsfig,subfigure}
\definecolor{myred}{cmyk}{0.000000,1.000000,1.000000,0.1}
\definecolor{myblue}{rgb}{0.000000,0.000000,1.000000}

\newtheorem*{remark}{Remark}

\usepackage{natbib}
\usepackage[allcolors=myblue]{hyperref}

\usepackage{tikz}

\usepackage{cite}
\newcommand{\BE}{\begin{equation}}
	\newcommand{\EE}{\end{equation}}
\newcommand{\BEstar}{\begin{equation*}}
	\newcommand{\EEstar}{\end{equation*}}

\graphicspath{{figures/}}

\usepackage[T1]{fontenc}

\usepackage{aecompl}

\date{}
\journal{a journal}

\begin{document}
\begin{frontmatter}
\title{CD-GAN: a robust fusion-based generative adversarial network for unsupervised remote sensing change detection with heterogeneous sensors}

\tnotetext[]{Part of this work has been supported by the ANR-3IA Artificial, Natural Intelligence Toulouse Institute (ANITI) under grant agreement ANITI ANR-19-P3IA-0004, NSFC (Grant No. 12171072), Key Projects of Applied Basic Research in Sichuan Province (Grant No. 2020YJ0216), and National Key Research and Development Program of China (Grant No. 2020YFA0714001).}

\author[SMU]{Jin-Ju Wang\corref{cor1}} 
\ead{jinjuwang123@163.com}
\author[n7]{Nicolas Dobigeon}
\ead{nicolas.dobigeon@enseeiht.fr}
\author[n7]{Marie Chabert}
\ead{marie.chabert@enseeiht.fr}
\author[UESTC]{Ding-Cheng Wang}
\ead{wangdc@uestc.edu.cn}
\author[UESTC]{Ting-Zhu Huang}
\ead{tingzhuhuang@126.com}
\author[UESTC]{Jie Huang}
\ead{huangjie\_uestc@uestc.edu.cn}
\address[SMU]{School of Mathematics, Southwest Minzu University, Chengdu, Sichuan 610041, China}
\address[n7]{University of Toulouse, IRIT/INP-ENSEEIHT Toulouse, BP 7122, 31071 Toulouse Cedex 7, France}
\address[UESTC]{School of Mathematical Sciences, University of Electronic Science and Technology of China (UESTC), Chengdu, Sichuan, China}
\cortext[cor1]{Corresponding author}

\begin{abstract}
{In the context of Earth observation, change detection boils down to  comparing  images acquired at different times by sensors of possibly different spatial and/or spectral resolutions or  different modalities (e.g., optical or radar).
Even when considering only optical images, this task has proven to be challenging as soon as the sensors differ by their spatial and/or spectral resolutions.
This paper proposes a novel unsupervised change detection method dedicated to images acquired by such so-called heterogeneous optical sensors. It capitalizes on recent advances which formulate the change detection task into a robust fusion framework. Adopting this formulation, the work reported in this paper shows that any off-the-shelf network trained beforehand to fuse optical images of different spatial and/or spectral resolutions can be easily complemented with a network of the same architecture and embedded into an adversarial framework to perform change detection. A comparison with state-of-the-art change detection methods demonstrates the versatility and the effectiveness of the proposed approach.}
\end{abstract}

\begin{keyword}
heterogeneous sensors\sep change detection\sep adversarial training \sep fusion \sep spatial sparsity. 
\end{keyword}
\end{frontmatter}

\section{Introduction}
In the context of Earth observation, change detection (CD) aims at identifying spatial areas that have been altered between two remote sensing image acquisitions \citep{tewkesbury2015critical, bruzzone2012novel}. 
With the rapid development of urban areas and the frequent occurrence of natural disasters like floods and earthquakes, CD has become an essential task to monitor land cover evolution
\citep{gueguen2016toward, hegazy2015monitoring, wang2010comparison, manonmani2010remote}.
To be effective, {the CD methods have to account for the intrinsic characteristics of the two sensors acquiring the pair of images and their  possible dissimilarities.
When the sensors share the same modality and have the same spatial and spectral resolutions, the images they provide are usually referred to as homogeneous. Conversely, when the sensors differ by their modality or resolutions, the images are referred to as heterogeneous.}

Due to the rapid development of remote sensing technologies, there is a growing availability of images coming from heterogeneous sensors. 
This offers new opportunities to perform timely CD, which may be valuable in emergency situations. In this context, this paper specifically focuses on the challenging CD applicative scenario that involves optical sensors with different spatial and/or spectral resolutions. An archetypal scenario, considered in this paper for illustrative purpose and generally referred to as \emph{complementary acquisition}, involves two optical images of complementary resolutions, i.e., a high spatial and low spectral resolution (HRLS) image and a low spatial and high spectral resolution (LRHS) image. Adopting a conventional dichotomy over the optical sensors, this could be \emph{i) }a pair composed of a panchromatic (PAN) image and a multispectral (MS) image, \emph{ii)} a pair composed of a PAN image and a hyperspectral (HS) image or \emph{iii)} a pair composed of a MS image and an HS image.	

{The pipeline of most CD methods consists of three steps: image pre-processing, change image (CI) generation and binary change map (CM) generation \citep{bruzzone2002adaptive}. 
In particular, when dealing with heterogeneous sensors, the pre-processing step aims at allowing a reliable pixel-by-pixel comparison to be performed between the two acquired images. The second step then generates the CI by comparing the pre-processed images pixel-by-pixel. 
The last step performs high level processing such as segmentation or classification on the CI to label the altered pixels and/or group of pixels. 
 
When facing dissimilar spatial and/or spectral resolutions, the first step of the CD pipeline outlined above mainly rely on individual processing applied to each acquired image. It produces a pair of virtually observed images with the same spatial and spectral resolutions where two homologous pixels (e.g., with the same coordinates) depict the same spatial location in the observed scene. This step can be achieved following various strategies, such as filtering, interpolation, resampling, deconvolution  or super-resolution \citep{li2017change, gao2016automatic, niu2018conditional, zhao2017discriminative, li2021spatially, alberga2007performance, alberga2007comparison}. 
For instance, a virtual HRLS image can be obtained by spatially interpolating and spectrally filtering an observed LRHS image. 
Another possibility is to spatially downsample the observed HRLS image and spectrally filtering the LRHS image to produce a pair of virtual images with the same low spatial and low spectral resolutions (LRLS). 
Thereby, the two acquired images are  embedded into the same feature space where their virtual counterparts can be compared pixel-by-pixel \citep{nielsen1998multivariate, nielsen2007regularized}. Instead of applying these transformations in the image domain, one alternative may consist in deriving a patch-based graph representation of each image. The structures of the two resulting graphs are then compared to locate changes \citep{sun2020patch,sun2021structure,sun2021iterative}.

However, the above pre-processing strategies are implemented on each acquired image individually. Thus they do not fully exploit their interdependence and complementarity in terms of spatial and spectral information. Moreover, some of these pre-processing consist in artificially degrading the spatial and/or spectral resolutions of at least one of the acquired images, obviously resulting in a loss of information. Instead of individually pre-processing the two acquired images, a general CD framework, initially introduced in \citep{ferraris2017detecting} and later extended in \citep{Ferraris_IEEE_Trans_CI_2017,Ferraris_INFFUS_2020}, has envisioned the CD problem from the prism of multiband image fusion. Interestingly, this task shares some key characteristics with the problem of CD between heterogeneous but complementary images. It aims at recovering a latent (i.e., unobserved or virtual) image by merging the spatial and spectral information brought by a pair of observed images with complementary resolutions \citep{PSGAN,wang2019deep,dian2020regularizing,xu2020hyperspectral,zhang2020deep,wang2021enhanced,hu2021hyperspectral,li2023learning}. Tackling the CD task under the fusion paradigm, the spatial and spectral information of the two observed images is fully preserved, which enables the subsequent recovery of a CD map with the highest spatial and spectral resolutions. This strategy has demonstrated its superiority over state-of-the-art CD methods that mostly rely on feature space embedding.

Meanwhile, some recent works exploit the versatility of deep neural networks to perform CD in case of homogeneous or heterogeneous sensors \citep{shafique2022deep,liu2016deep, zhao2017discriminative,liu2019bipartite,zhang2021escnet}. 
For instance, the so-called slow feature analysis proposed by \citet{du2019unsupervised} consists in comparing the features learnt by two deep networks, each dedicated to one of the observed image. Again, this strategy does not fully exploit the information shared by the two images since the features are extracted from the two images individually. In \citep{Cycle-GAN}, cycle-consistent adversarial networks (CycleGANs) learn a subimage-to-subimage mapping, that allows the LRHS image to be embedded into a HRLS image space. This image translation can be also achieved thanks to particular network architectures embedding transformers to finally generate the CI \citep{xu2023ucdformer}. However, most of these deep learning-based CD methods face several major shortcomings. Firstly, very few consider the case of hetereogeneous acquisitions and most of the methods would need to be carefully adapted to handle images of different spatial and/or spectral resolutions. Secondly, the deep architectures solving the image translation problem  hardly preserve the whole spatial and spectral information brought by the pair of images, which may affect the objects and backgrounds. Thirdly, many of these deep architectures are primarily designed for supervised CD, i.e., their training requires image pairs taken before and after changes, with annotated changes. 

This paper proposes an unsupervised CD method for heterogeneous sensors (e.g., handling a HRLS image and a LRHS image) by capitalizing on the fusion-based CD framework mentioned above. This framework has the great advantage of jointly exploiting the spatial and spectral information brought by the two observed images, contrary to most of the approaches discussed above which apply individual pre-processing. Moreover it leverages the exceptional performance reached by recent neural fusion architectures, without requiring image pairs with annotated changes for training. More precisely, this paper shows that any off-the-shelf pre-trained deep network designed to fuse heterogeneous images can be reused as part of an adversarial architecture to perform unsupervised CD. This fusion network, whose design can be left to the final user, can be systematically complemented with a network of the same architecture to estimate a CI image. When compared to more conventional approaches, the strategy adopted in this paper appears to be more flexible while competing favorably. Moreover, the proposed framework can be kept up-to-date by embedding even more powerful pre-trained fusion networks, taking benefit from expected future advances in that domain.

The main contributions of this paper can be summarized as follows:  \emph{i)} after casting the heterogeneous CD problem into a robust fusion framework, we show that this framework can be interpreted as a generative adversarial model, whose generator is composed of two essential building blocks, \emph{ii)} since one of this block boils down to a fusion procedure, we suggest that any off-the-shelf fusion algorithm, such as a pre-trained deep neural network, can be easily complemented by a deep neural network to build an adversarial architecture dedicated to CD and \emph{iii)} exploiting the property of fusion consistency, a specific loss function is derived  to allow the overall adversarial architecture to be trained within an unsupervised generative framework.}

This article is organized as follows. 
Section \ref{sec:robust_fusion} recalls how the problem of CD can be cast as a robust fusion task. Capitalizing on this formulation, an adversarial framework is proposed in Section \ref{sec:framework} and a possible architecture is detailed for the particular case of complementary acquisitions. Experimental results obtained on simulated datasets are reported in Section \ref{sec:experiments_synth} to assess the efficiency, the versatility and the robustness of the proposed approach when compared to alternative CD methods. Section \ref{sec:experiments_real} illustrates the approach relevance when analyzing real datasets.
Finally, Section \ref{sec:conclusion} concludes the paper.

\section{Background: robust fusion-based change detection}\label{sec:robust_fusion}
This paper focuses on the problem of detecting changes between two optical images denoted $\mathbf{Y}_{1} \in \mathbb{R}^{m_{1}\times n_{1}} $ and $\mathbf{Y}_{2} \in \mathbb{R}^{m_{2}\times n_{2}}$ acquired over the same scene, at different times $t_{1}$ and $t_{2}$ by two sensors $\textsf{S}_1$ and $\textsf{S}_2$, respectively. 
{Here, $m_{i}$ and $n_{i}$ stand for the numbers of bands and pixels of the image $\mathbf{Y}_{i}$ ($i\in \left\{1,2\right\}$), respectively. }
In our case of interest, these two images are assumed to have different spatial and/or spectral resolutions, i.e., $n_{1}\neq n_2$ and/or $m_{1}\neq m_2$. 

\begin{remark}[Complementary acquisitions]
Throughout this paper, we will consider the particular case of images with complementary resolutions, as in \citep{Ferraris_IEEE_Trans_CI_2017}. This choice is made for illustrative purposes and does not sacrifice the generality of the approach. This case can be stated by the following relations
\begin{equation}
m_{1}\geq m_{2} \ \text{and} \ 
n_{2}\geq n_{1}. \label{eq:dim}
\end{equation}
In other words, $\mathbf{Y}_{1}$ and $\mathbf{Y}_{2}$ are LRHS and HRLS images respectively. 
\end{remark}

{Because of the different spatial and/or spectral resolutions between $\mathbf{Y}_{1}$ and $\mathbf{Y}_{2}$, a naive pixel-wise comparison of the two images cannot be conducted to locate the changes. To overcome this difficulty, the adversarial strategy proposed in this paper builds upon the robust fusion framework proposed in \citep{Ferraris_IEEE_Trans_CI_2017} and generalized in \citep{Ferraris_INFFUS_2020}. This fusion framework relates the two observed images $\mathbf{Y}_{1}$ and $\mathbf{Y}_{2}$ and two latent images $\mathbf{X}_{1} \in \mathbb{R}^{m\times n}$ and $\mathbf{X}_{2} \in \mathbb{R}^{m\times n}$ which share the same high spatial and high spectral resolutions with $n \geq \max \left\{n_1,n_2\right\}$ and $m \geq \max\left\{m_1,m_2\right\}$. }
More precisely, the latent and acquired images are related through  the observation models
\begin{subequations}\label{eq:directmodels}
\begin{align}
    \mathbf{Y}_{1} &= \mathcal{H}_1\left(\mathbf{X}_{1}\right)\\
    \mathbf{Y}_{2} &= \mathcal{H}_2\left(\mathbf{X}_{2}\right)
\end{align}
\end{subequations}
where $\mathcal{H}_1: \mathbb{R}^{m\times n}  \rightarrow \mathbb{R}^{m_{1}\times n_{1}} $ and $\mathcal{H}_2 : \mathbb{R}^{m\times n}  \rightarrow \mathbb{R}^{m_{2}\times n_{2}}$ stand for degradation operators. These operators are assumed to be linear, i.e., $\mathcal{H}_i(\mathbf{A}+\lambda\mathbf{B}) = \mathcal{H}_i(\mathbf{A})+\lambda\mathcal{H}_i(\mathbf{B})$ for $i\in \left\{1,2\right\}$, $\mathbf{A}\in \mathbb{R}^{m\times n}$, $\mathbf{B}\in \mathbb{R}^{m\times n}$ and $\lambda \in \mathbb{R}$. 

\begin{remark}[Complementary acquisitions]
In case of complementary acquisitions, the operators $\mathcal{H}_1\left(\cdot\right)$ and $\mathcal{H}_2(\cdot)$ stand for spatial and spectral degradations respectively. 
As in \citep{Ferraris_IEEE_Trans_CI_2017} and in most of previous works from the literature dealing with multiband imaging fusion \citep{wei2015fast,dian2020regularizing,xu2020hyperspectral,wang2021enhanced,hu2021hyperspectral}, they are defined as
\begin{align*}
     \mathcal{H}_1\left(\mathbf{X}_1\right) &\approx \mathbf{X}_1\mathbf{R},\\
      \mathcal{H}_2\left(\mathbf{X}_2\right) &\approx \mathbf{L}\mathbf{X}_2.
\end{align*}
Here, the matrix $\mathbf{R}$ is generally decomposed into $\mathbf{R}=\mathbf{BS}$, where $\mathbf{B}$ stands for a spatially-invariant blurring operation and $\mathbf{S}$ corresponds to a regular subsampling. 
The matrix $\mathbf{L}$ is a spectral degradation specified by the sensor spectral response. 
The symbol $\approx$ accounts for potential mismodeling or measurement noise. {The latent space $\mathbb{R}^{m\times n}$ embeds images with spectral and spatial resolutions defined by the LRHS and HRLS images respectively, i.e., $m=m_1$ and $n=n_2$.}
\end{remark}

{It is worth noting that the latent images $\mathbf{X}_{1}$ and $\mathbf{X}_{2}$ share the same spatial and spectral resolutions. As a consequence, these two latent images could be pixel-wisely compared to detect any changes that may have occurred between the two acquisition times. Based on this finding, CD can be reframed as an inference problem referred to as robust fusion. The resolution of this problem can be decomposed into two main steps, which are detailed below.}

\subsection{CI inference} \label{subsubsec:CIstep} 
Recall that the two latent images $\mathbf{X}_{1}$ and $\mathbf{X}_{2}$ share the same spatial and spectral resolutions. If they were available, a CI denoted $\Delta \mathbf{X} \in  \mathbb{R}^{m\times n}$ of high spatial and high spectral resolutions could be easily derived by a pixel-by-pixel difference. More precisely, this change image is computed as
\begin{equation}
\label{hrchange}
    \Delta \mathbf{X} = \mathbf{X}_{2} - \mathbf{X}_{1}
\end{equation}
{with $\Delta\mathbf{X}=\left[\Delta\mathbf{x}_1,\ldots,\Delta    \mathbf{x}_n\right]$, where $\Delta\mathbf{x}_i$ is the $m \times 1$ change vector associated to the pixel $i \in    \{1,...,n\}$. 

Then, conventional CD methods dedicated to homogeneous sensors, such as change vector analysis (CVA) and its extensions  \citep{bovolo2006theoretical, du2020improved}, can  be used to derive the binary CM from the CI.} In its canonical formulation, CVA consists in computing the pixel-wise energy $\mathbf{e}=\left[{e}_1,\ldots,{e}_n\right] \in \mathbb{R}^{n}$ of the change image where, for $i = 1, \ldots, n$,
\begin{equation} \notag
e_{i}=\lVert \Delta\mathbf{x}_{i}\rVert_{2}.
\end{equation}
The components $e_{i}$ of $\mathbf{e}$ with the smallest (resp. highest) values most likely correspond to unchanged (resp. changed) pixels, whose indices are gathered in the set $\Omega_{\textrm{u}}$ (resp. $\Omega_{\textrm{c}}$). Thus, a natural decision rule consists in thresholding the energy change image, i.e., for $i = 1, \ldots, n$
\begin{equation}
i\in\left\{
\begin{array}{cc}
		 \Omega_{\textrm{c}} & \text{if $e_{i} > \tau$}, \\
		 \Omega_{\textrm{u}}& \text{otherwise}.
\end{array}\right.
\end{equation}
{Here, the threshold $\tau$ adjusts the trade-off between the probability of false alarm and the probability of detection. It can be chosen using a dedicated method, as in \citep{ostu}.} Then, the final binary CM $\mathbf{d} =[d_{1},\cdots, d_{n}] \in \{0, 1\}^{n}$ can be derived as 
\begin{equation}
d_{i}=\left\{\begin{array}{cc}
	1 & \text{if $i\in \Omega_{\textrm{c}}$}, \\
	0  & \text{if $i\in \Omega_{\textrm{u}}$}. 
\end{array}\right.
\end{equation}
{It is worth noting that, as stated in many works such as \citep{sun2021structure}, the changes are generally related to any evolution of the land cover, e.g., due to natural disasters, seasonal trends or human induced modifications. Conversely, when dealing with optical images, variations due to illumination should not be identified as changes since they result from the acquisition conditions. Under this paradigm, the conventional formulation of CD assumes that most of the pixels are expected to be unaffected by changes. In other words, CD is a particular instance of an unbalanced binary classification problem since it consists in recovering the binary CM $\mathbf{d}$ with $\mathrm{card}\{\Omega_{\textrm{u}}\} \gg \mathrm{card}\{\Omega_{\textrm{c}}\}$. Thus, the threshold $\tau$ should be adjusted to identify the most relevant changes in the CI, in particular those that cannot be explained only by illumination variations.}

\subsection{Fusion} \label{subsec:fusionstep} Reciprocally, if the CI $\Delta\mathbf{X}$ was known, the latent images $\mathbf{X}_1$ and $\mathbf{X}_2$ could be inferred by solving a multiband image fusion problem. Indeed, adopting the framework introduced in \citep{Ferraris_IEEE_Trans_CI_2017} and exploiting the identity \eqref{hrchange} and the linearity of the degradation operators, the direct models  \eqref{eq:directmodels}  can be equivalently rewritten as
\begin{subequations}\label{eq:directmodels2}
\begin{align}
         {\mathbf{Y}}_{1} &= \mathcal{H}_1\left(\mathbf{X}_{1}\right),  \label{eq:directmodels2_Y1}\\
    \widetilde{\mathbf{Y}}_{2} &= \mathcal{H}_2\left(\mathbf{X}_{1}\right)\label{eq:directmodels2_Y2}
\end{align}
\end{subequations}
where\footnote{In this paper, we introduced the corrected image at time $t_1$. 
Another possible choice would consist in defining the corrected image $\widetilde{\mathbf{Y}}_{1}\triangleq {\mathbf{Y}}_{1}+\mathcal{H}_1\left(\Delta\mathbf{X}\right)$ that would be acquired by the sensor $\textsf{S}_1$ at time $t_2$, i.e., after the changes occur. In this case, the subsequent technical derivations should be adapted accordingly.}
\begin{equation}\label{eq:Y2tilde}
    \widetilde{\mathbf{Y}}_{2}\triangleq {\mathbf{Y}}_{2}-\mathcal{H}_2\left(\Delta\mathbf{X}\right),
\end{equation}
is the so-called \emph{corrected} image, i.e.,  the image that would be acquired by $\textsf{S}_2$ at $t_1$ (before a change occurs). 
From the set of equations \eqref{eq:directmodels2}, it clearly appears that recovering the latent image  $\mathbf{X}_{1}$ from the observed image ${\mathbf{Y}}_{1}$ and the corrected image $\widetilde{\mathbf{Y}}_{2}$ is an image fusion task. 
The other latent image $\mathbf{X}_{2}$ can be subsequently derived from $\mathbf{X}_{1}$ and $\Delta\mathbf{X}$ using \eqref{hrchange}.

\begin{remark}[Complementary acquisitions]
{Fusing a pair of images resulting from complementary acquisitions aims at recovering a latent image whose spatial and spectral resolutions are the highest of the observed image ones. This problem
has motivated a wide bunch of research works over the three last decades, specifically dealing with pansharpening when fusing PAN and MS images \citep{vivone2014critical}, hyperspectral pansharpening when fusing PAN and HS images \citep{Loncan_IEEE_GRS_Mag_2015,dian2023zero} or multiband image fusion when fusing MS and HS images \citep{Wei_IEEE_JSTSP_2015,dian2020regularizing,li2023learning}.}
\end{remark}

\subsection{From fusion consistency to CD}\label{subsec:consistency}
{From what precedes, heterogeneous CD can be decomposed into two entangled steps, namely CI inference and fusion. These two steps cannot be performed independently since each step relies on key quantities provided as outputs by the other step. To overcome this difficulty, the robust fusion-based CD approach leverages on a consistency property fulfilled by any fusion method.} This property states that a well-designed fusion process should be reversible. 
This means that the observed image $\mathbf{Y}_1$ is expected to be as close as possible to the image that would be obtained by applying the acquisition process to the estimated fused image $\hat{\mathbf{X}}_1$ \citep{wald1997fusion}. 
In other words, the so-called \emph{predicted} image defined by 
\begin{equation}\label{eq:Y1hat}
  \hat{\mathbf{Y}}_1 = \mathcal{H}_1(\hat{\mathbf{X}}_1).
\end{equation}
 that would be observed by the sensor $\textsf{S}_1$ at time $t_1$ is expected to be practically indistinguishable from the observed image $\mathbf{Y}_1$. This consistency property is expressed as
\begin{equation} \label{eq:consistency_Y1}
  \hat{\mathbf{Y}}_1 \approx \mathbf{Y}_1.
\end{equation}
{It is the main rationale of the strategy introduced in \citep{ferraris2017detecting}, where CD is conducted by alternating between the CM inference and the fusion steps. 
These two steps are achieved by iteratively solving two optimization problems whose complexities depend on the acquisition scenario. 
In the next section, we show that the robust-fusion based CD approach can be envisioned from an adversarial paradigm, by jointly conducting CM inference and fusion in a unified framework. 
In particular, this framework embeds two essential building blocks dedicated to the two aforementioned steps. Notably, since one of these tasks consists in solving a fusion problem, any off-the-shelf pre-trained network designed for this purpose can be effectively reused as an essential building block. This framework is detailed in what follows.}

\section{Proposed adversarial framework}\label{sec:framework}
\begin{figure*}
\centering
    \includegraphics[width=\linewidth]{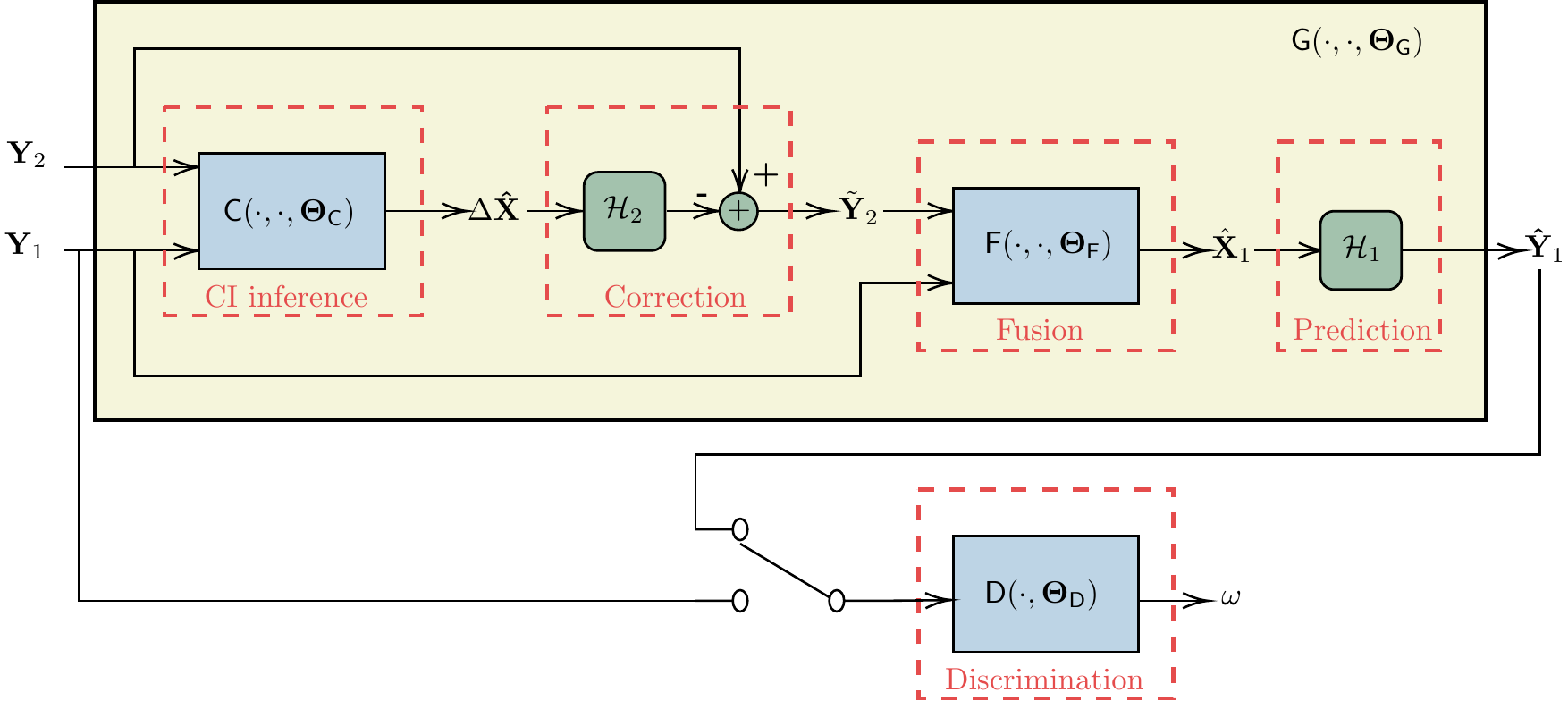}
    \caption{Flowchart of the overall proposed CD-GAN architecture. The 3 main networks $\mathsf{C}\left(\cdot,\cdot;\boldsymbol{\Theta}_{\mathsf{C}}\right)$, $\mathsf{F}\left(\cdot,\cdot;\boldsymbol{\Theta}_{\mathsf{F}}\right)$ and $\mathsf{D}\left(\cdot,\cdot;\boldsymbol{\Theta}_{\mathsf{D}}\right)$ appear as blue light boxes. The virtual generator $\mathsf{G}\left(\cdot,\cdot;\boldsymbol{\Theta}_{\mathsf{G}}\right)$ is figured as a beige  box. {The five main sub-tasks of the architecture are highlighted as red  dashed-lined boxes.}}
    \label{flowchart}
\end{figure*}

The previous section showed that CD between images acquired by heterogeneous sensors can be conducted by inferring a  latent image pair $\left\{\mathbf{X}_1, \mathbf{X}_2\right\}$, or one latent image and the CI $\left\{\mathbf{X}_{1}, \Delta\mathbf{X}\right\}$ with $\mathbf{X}_2=\mathbf{X}_1+\Delta\mathbf{X}$. 
{This inference problem, coined as robust fusion, relies on two steps, namely CI inference and fusion, detailed in Sections \ref{subsubsec:CIstep} and \ref{subsec:fusionstep}, respectively. 
This paper shows that these two steps can be performed by two essential building blocks embedded into an unifying adversarial architecture.} 
This architecture offers a high degree of modularity and thus opens the door to countless refinement opportunities and subsequent performance improvements. The different building blocks of the generative adversarial architecture, depicted in Fig. \ref{flowchart} and referred to as CD-GAN in the sequel, as well as the training strategy are detailed in this section. 
Finally, we provide a specific instance of this framework for the particular case of complementary acquisitions.

\subsection{Overall architecture}

\noindent \textbf{CI inference -- } Estimating the CI $\Delta{\mathbf{X}}$ from the pair of observed images $\left(\mathbf{Y}_{1},\mathbf{Y}_{2}\right)$ can be interpreted as a mapping from the product set $\mathbb{R}^{m_1 \times n_1} \times \mathbb{R}^{m_2 \times n_2}$ of observed images towards the set $\mathbb{R}^{n \times m}$ of CIs. 
In the proposed framework, this unknown mapping is defined as 
\begin{equation}
    {\Delta\hat{\mathbf{X}}} = \mathsf{C}\left(\mathbf{Y}_{1},\mathbf{Y}_{2};\boldsymbol{\Theta}_{\mathsf{C}}\right) \label{eq:GANmodel_C}
\end{equation}
where $\mathsf{C}\left(\cdot,\cdot;\boldsymbol{\Theta}_{\mathsf{C}}\right)$ is a deep network parameterized by $\boldsymbol{\Theta}_{\mathsf{C}}$. 
This network is able to solve the CD problem in the case of heterogeneous sensors. 
{Indeed, once trained, it directly provides the estimated CI $\Delta\hat{\mathbf{X}}$, from which the binary change map $\hat{\mathbf{d}}$ can be estimated as detailed in Section \ref{subsubsec:CIstep}. 
In a supervised setting, training this network would require triplets $\left(\mathbf{Y}_1^{(i)}, \mathbf{Y}_2^{(i)}, \Delta\mathbf{X}_1^{(i)}\right)$ composed of a pair of observed images and the corresponding change image, which may significantly limit the applicability of the method. 
Conversely, the proposed adversarial approach will propose to train this network within an unsupervised setting, i.e., without requiring the availability of such triplets.}\\

\noindent \textbf{Fusion -- } Using the notations introduced in the previous section, the fusion step can be formulated as a mapping from the product set $\mathbb{R}^{m_1 \times n_1} \times \mathbb{R}^{m_2 \times n_2}$ of observed images to the set $\mathbb{R}^{n \times m}$ of latent  images, i.e., 
\begin{equation}
    \hat{\mathbf{X}}_{1} =  \mathsf{F}({\mathbf{Y}}_{1},\tilde{\mathbf{Y}}_{2};\boldsymbol{\Theta}_{\mathsf{F}}), \label{eq:GANmodel_F}
\end{equation}
where $\mathsf{F}\left(\cdot,\cdot;\boldsymbol{\Theta}_{\mathsf{F}}\right)$ stands for a fusion method parameterized by $\boldsymbol{\Theta}_{\mathsf{F}}$. 
{In this work, we propose to benefit from recent efforts promoting deep architectures specifically designed to solve the fusion task \citep{wang2019deep,dian2020regularizing,wang2021enhanced,SSRNET}.
More precisely, in the sequel of the paper, we will assume that the fusion method $\mathsf{F}\left(\cdot,\cdot;\boldsymbol{\Theta}_{\mathsf{F}}\right)$ can be chosen as any off-the-shelf pre-trained deep network. Thus  $\boldsymbol{\Theta}_{\mathsf{F}}$ remains fixed in the following. 
It is worth noting that this fusion network implicitly requires the knowledge of the CI $\Delta\mathbf{X}$ (or one estimate $\Delta\hat{\mathbf{X}}$), which is provided by the network $\mathsf{C}\left(\cdot,\cdot;\boldsymbol{\Theta}_{\mathsf{C}}\right)$. Indeed, this network takes as one input the corrected image $\tilde{\mathbf{Y}}_2$ which depends on the CI $\Delta\mathbf{X}$ according to  \eqref{eq:Y2tilde}.}\\

\noindent  \textbf{Fusion consistency-based training -- } {A naive strategy to train the network $\mathsf{C}\left(\cdot,\cdot;\boldsymbol{\Theta}_{\mathsf{C}}\right)$ dedicated to CI inference would consist in ensuring that the output $\hat{\mathbf{X}}_1$ of the subsequent correction and fusion steps defined by \eqref{eq:GANmodel_F} closely matches the true latent image ${\mathbf{X}}_1$. This approach, generally adopted by the GAN-based fusion methods from the literature, relies on an appropriate discriminator, denoted as $\mathsf{D}\left(\cdot;\boldsymbol{\Theta}_{\mathsf{D}}\right)$ in Fig. \ref{flowchart}. It would require the availability of a training set  of triplets $\left(\mathbf{Y}_1^{(i)}, \mathbf{Y}_2^{(i)}, \mathbf{X}_1^{(i)}\right)$ composed of the two observed images and the associated latent image standing for the expected fusion result. Such a supervised learning appears restrictive since, in practice, it requires the availability of representative labelled data. This work proposes an alternative training that only requires the observed image $\mathbf{Y}_1$. More precisely, by leveraging the fusion consistency \eqref{eq:consistency_Y1} discussed in Section \ref{subsec:consistency},  one designs a discriminative network which assesses the quality of the predicted image $\hat{\mathbf{Y}}_1$ defined by \eqref{eq:Y1hat} with respect to (w.r.t.) the observed image $\mathbf{Y}_1$. This can be achieved by designing a discriminative network $\mathsf{D}(\cdot;\boldsymbol{\Theta}_{\mathsf{D}})$ such that }
\begin{equation}
  \omega = \mathsf{D}(\boldsymbol{\Upsilon};\boldsymbol{\Theta}_{\mathsf{D}})
\end{equation}
where $\boldsymbol{\Upsilon}\in\{{\mathbf{Y}}_1, \hat{\mathbf{Y}}_1\}$, $\omega \in \left\{0,1\right\}$ is the binary label stating the likeness between the predicted image  $\hat{\mathbf{Y}}_1$ and the observed image ${\mathbf{Y}}_1$ ($\omega=1$ meaning ``alike'' and $\omega=0$ meaning ``different''), and $\boldsymbol{\Theta}_{\mathsf{D}}$ is the set of the discriminative network parameters.\\

{To summarize, the overall proposed adversarial architecture can be decomposed into five main sub-tasks depicted as red  dashed-lined boxes in Fig. \ref{flowchart} and listed below:}
\begin{itemize}
    \item \textit{CI inference:} the generative network $\mathsf{C}\left(\cdot,\cdot;\boldsymbol{\Theta}_{\mathsf{C}}\right)$ produces a candidate CI ${\Delta\hat{\mathbf{X}}}$ from the two observed images ${\mathbf{Y}}_1$ and ${\mathbf{Y}}_2$,
    \item \textit{Correction:} the infered candidate CI ${\Delta\hat{\mathbf{X}}}$ is used to produce the corrected image $\tilde{\mathbf{Y}}_{2}$  following \eqref{eq:Y2tilde},
    \item \textit{Fusion:} the corrected image  $\tilde{\mathbf{Y}}_{2}$ is subsequently fused with the observed image ${\mathbf{Y}}_{1}$ thanks to the pre-trained network $\mathsf{F}\left(\cdot,\cdot;\boldsymbol{\Theta}_{\mathsf{F}}\right)$, providing the estimated latent image $\hat{\mathbf{X}}_1$,
    \item \textit{Prediction:} the predicted image $\hat{\mathbf{Y}}_1=\mathcal{H}_1(\hat{\mathbf{X}}_1)$ is derived,
    \item \textit{Discrimination:} the quality of the predicted image $\hat{\mathbf{Y}}_1$ w.r.t. the observed image ${\mathbf{Y}}_1$ is assessed thanks to the discriminative network $\mathsf{D}(\cdot;\boldsymbol{\Theta}_{\mathsf{D}})$.
\end{itemize}
{A meaningful interpretation of the proposed architecture can be drawn by observing that the sub-network highlighted as a beige box in Fig. \ref{flowchart} and denoted $\mathsf{G}(\cdot,\cdot;\boldsymbol{\Theta}_{\mathsf{G}})$ acts as a generator with parameters $\boldsymbol{\Theta}_{\mathsf{G}} = \left\{\boldsymbol{\Theta}_{\mathsf{C}},\boldsymbol{\Theta}_{\mathsf{F}}\right\}$. It takes the two observed images $\mathbf{Y}_1$ and $\mathbf{Y}_2$ as inputs and embeds the four first sub-tasks of the architecture to produce the predicted image $\hat{\mathbf{Y}}_1$. This can be written as}
\begin{equation}
    \mathsf{G}(\mathbf{Y}_1,\mathbf{Y}_2;\boldsymbol{\Theta}_{\mathsf{G}}) \triangleq  \hat{\mathbf{Y}}_1,\label{eq:generator_prediction}
\end{equation}
where we recall that
\begin{equation}\notag
\begin{aligned}
\hat{\mathbf{Y}}_1 &= \mathcal{H}(\hat{\mathbf{X}}_1)\\
\hat{\mathbf{X}}_1 &= \mathsf{F}({\mathbf{Y}}_{1},\tilde{\mathbf{Y}}_{2};\boldsymbol{\Theta}_{\mathsf{F}})\\
\tilde{\mathbf{Y}}_{2} &= {\mathbf{Y}}_{2}-\mathcal{H}_2(\Delta\hat{\mathbf{X}})\\ 
\Delta\hat{\mathbf{X}} &= \mathsf{C}\left(\mathbf{Y}_{1},\mathbf{Y}_{2};\boldsymbol{\Theta}_{\mathsf{C}}\right)
\end{aligned}
\end{equation}
{or, compactly written,}
\begin{equation*}
    {\mathsf{G}(\mathbf{Y}_1,\mathbf{Y}_2;\boldsymbol{\Theta}_{\mathsf{G}}) = \mathcal{H}\left(\mathsf{F}\left({\mathbf{Y}}_{1}, {\mathbf{Y}}_{2}-\mathcal{H}_2\left(\mathsf{C}\left(\mathbf{Y}_{1},\mathbf{Y}_{2};\boldsymbol{\Theta}_{\mathsf{C}}\right)\right);\boldsymbol{\Theta}_{\mathsf{F}}\right)\right).}
\end{equation*}
{Since the fusion network is assumed to have been pre-trained, the sole generator parameters that still require training are the parameters $\boldsymbol{\Theta}_{\mathsf{C}}$ of the CI inference network. Thus, in what follows, this generator will be denoted $\mathsf{G}(\cdot,\cdot;\boldsymbol{\Theta}_{\mathsf{C}})$. Introducing this generator allows the proposed architecture to be framed into an (almost) conventional GAN framework}\footnote{Another possible interpretation consists in pipelining the four last sub-tasks of the architecture to define an extended discriminator combined with a generator defined as the sole network dedicated to the CI inference}.

\subsection{Loss function}
As previously stated, the fusion network $\mathsf{F}\left(\cdot,\cdot;\boldsymbol{\Theta}_{\mathsf{F}}\right)$ can be chosen as any state-of-the-art network (or even any model-based fusion algorithm) from the literature and it is assumed to have been pre-trained (or calibrated) beforehand. 
{In this framework, two sub-networks need to be trained, namely the CI inference network $\mathsf{C}\left(\cdot,\cdot;\boldsymbol{\Theta}_{\mathsf{C}}\right)$ and the discriminative network $\mathsf{D}(\cdot;\boldsymbol{\Theta}_{\mathsf{D}})$, by minimizing a well-chosen loss function.}
First of all, the proposed loss function incorporates an adversarial cost defined by
\begin{multline}\label{eq:advloss}
\mathcal{L}_{\mathrm{adv}}\left(\boldsymbol{\Theta}_{\mathsf{C}},\boldsymbol{\Theta}_{\mathsf{D}}\right) = \mathbb{E}_{\mathbf{Y}_1}\left[\log \mathsf{D}\left({\mathbf{Y}}_1;\boldsymbol{\Theta}_{\mathsf{D}}\right)\right] \\+ \mathbb{E}_{\mathbf{Y}_1,\mathbf{Y_2}}\left[\log(1-\mathsf{D}(\mathsf{G}(\mathbf{Y}_1,\mathbf{Y}_2;\boldsymbol{\Theta}_{\mathsf{C}});\boldsymbol{\Theta}_{\mathsf{D}}))\right].
\end{multline}
{Then, as emphasized by previous works \citep{isola2017image}, the adversarial cost can benefit from its combination with application-oriented costs. The first one is the so-called prediction loss introduced to assess the quality of the predicted image $\hat{\mathbf{Y}}_1$ w.r.t. the observed image $\mathbf{Y}_1$}
\begin{equation}\label{eq:preloss}
    \mathcal{L}_{\mathrm{pre}}\left(\boldsymbol{\Theta}_{\mathsf{C}}\right) = \mathbb{E}_{\mathbf{Y}_1,\mathbf{Y_2}}\left[\|\mathbf{Y}_1 - \hat{\mathbf{Y}}_1\|_\textrm{F}^2\right]
\end{equation}
where $\|\cdot\|_\textrm{F}$ denotes the Frobenius norm. {Since the changes are expected to affect only a few pixels, as discussed in Section \ref{subsubsec:CIstep}, a second term promoting spatial sparsity of the estimated CI is also introduced \citep{Ferraris_IEEE_Trans_CI_2017,Ferraris_INFFUS_2020}}
\begin{equation}
    \mathcal{L}_{\mathrm{spa}}\left(\boldsymbol{\Theta}_{\mathsf{C}}\right) = \mathbb{E}_{\mathbf{Y}_1,\mathbf{Y}_2}\left[\|\Delta\hat{\mathbf{X}}\|_{2,1}\right]
\end{equation}
where $\|\cdot\|_{2,1}$ denotes a group sparsity promoting norm. {Finally, training both networks is formulated as the minimax problem}
\begin{equation}\label{eq:overallloss}
    \min_{\boldsymbol{\Theta}_{\mathsf{C}}} \max_{\boldsymbol{\Theta}_{\mathsf{D}}} \mathcal{L}_{\mathrm{adv}}\left(\boldsymbol{\Theta}_{\mathsf{C}},\boldsymbol{\Theta}_{\mathsf{D}}\right) + \alpha \mathcal{L}_{\mathrm{pre}}\left(\boldsymbol{\Theta}_{\mathsf{C}}\right) + \beta\mathcal{L}_{\mathrm{spa}}\left(\boldsymbol{\Theta}_{\mathsf{C}}\right),
\end{equation}
where $\alpha$ and $\beta$ are hyperparameters adjusting the relative weights of the terms. In practice, following the common training procedures of GAN-based architectures, the two networks are alternately updated by stochastic gradient-based optimization, i.e., the expectations involved in \eqref{eq:overallloss} are approximated by empirical averages over minibatchs.

\begin{remark}[Interpreting the generator training] Interestingly, the minimax optimization \eqref{eq:overallloss} underlying the generative-discriminative training can be interpreted from an image processing perspective. 
Indeed, let consider the deterministic setting or, equivalently, a minibatch composed of a unique sample. Under this simplifying assumption, training the generative network $\mathsf{G}\left(\cdot,\cdot;\boldsymbol{\Theta}_{\mathsf{C}}\right)$ dedicated to CI inference only boils down to solving the optimization problem
\begin{equation}
    \min_{\boldsymbol{\Theta}_{\mathsf{C}}} \|\mathbf{Y}_1 - \hat{\mathbf{Y}}_1\|_\textrm{F}^2 + \|\Delta\hat{\mathbf{X}}\|_{2,1} + \mathcal{L}_{\mathrm{adv}}\left(\boldsymbol{\Theta}_{\mathsf{C}}\right),
\end{equation}
where the predicted image $\hat{\mathbf{Y}}_1$ and the CI $\Delta\hat{\mathbf{X}}$ implicitly depend on $\boldsymbol{\Theta}_{\mathsf{C}}$. More precisely, since $\hat{\mathbf{Y}}_1=\mathcal{H}_1(\hat{\mathbf{X}}_1)$, the loss function can be rewritten as
\begin{equation} \label{eq:training_C}
    \min_{\boldsymbol{\Theta}_{\mathsf{C}}} \|\Delta \check{\mathbf{Y}}_1 - \mathcal{H}_1(\Delta{\hat{\mathbf{X}}})\|_\textrm{F}^2 + \|\Delta\hat{\mathbf{X}}\|_{2,1}+ \mathcal{L}_{\mathrm{adv}}\left(\boldsymbol{\Theta}_{\mathsf{C}}\right),
\end{equation}
where $\check{\mathbf{Y}}_1 \triangleq \mathcal{H}_1(\mathbf{X}_2)$ is the image that would be observed by the sensor $\textsf{S}_1$ at time $t_2$ and $\Delta \check{\mathbf{Y}}_1 \triangleq \check{\mathbf{Y}}_1 - {\mathbf{Y}}_1$ is the  $\textsf{S}_1$-virtual CI (i.e., the difference between the image actually observed at time $t_1$ by the sensor $\textsf{S}_1$ and the image virtually observed at time $t_2$ by the same sensor). Since the CI $\Delta\hat{\mathbf{X}}$ is fully defined as the output of the network $\mathsf{C}\left(\cdot,\cdot;\boldsymbol{\Theta}_{\mathsf{C}}\right)$ according to \eqref{eq:GANmodel_C}, training this network through the nonlinear optimization problem \eqref{eq:training_C} can be interpreted as a parametric inversion task: given the $\textsf{S}_1$-virtual CI $\Delta \check{\mathbf{Y}}_1$, one aims at inverting the measurement operator $\mathcal{H}_1(\cdot)$ seeking the solution in a parametric form specified by $\boldsymbol{\Theta}_{\mathsf{C}}$. It is worth noting that this inversion step has been explicitly referred to as a correction in \citep{Ferraris_IEEE_Trans_CI_2017,Ferraris_INFFUS_2020}. Note that in \eqref{eq:training_C}, we have introduced an additional task-driven data fitting term: the adversarial term $\mathcal{L}_{\mathrm{adv}}\left(\boldsymbol{\Theta}_{\mathsf{C}}\right)$.

\end{remark}

\subsection{Detailed architecture in case of complementary acquisitions}
\label{subsec:archi_complementary_acquisitions}
\begin{figure}[h!]
\centering
\includegraphics[width=\linewidth]{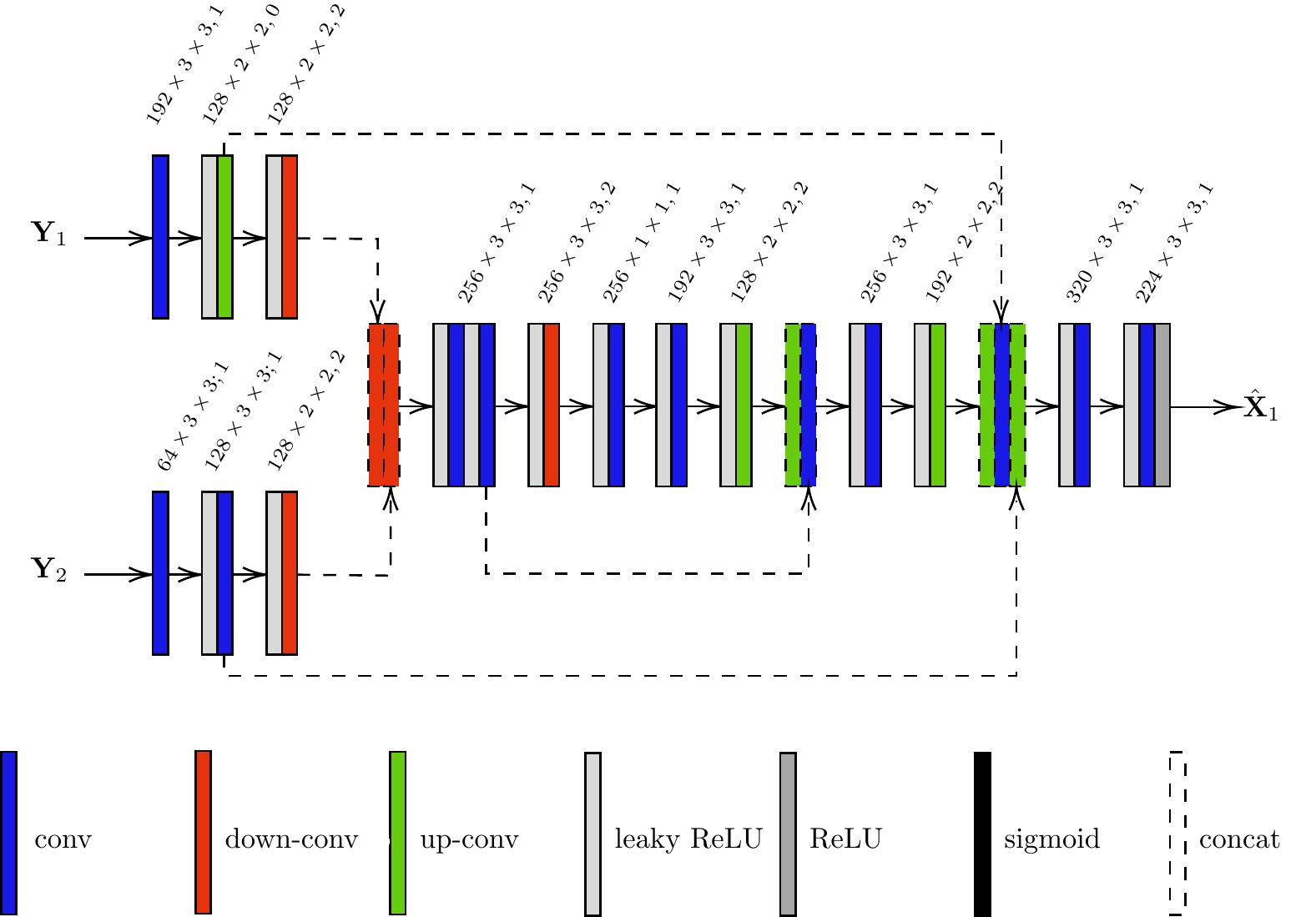}
    \caption{{Architecture of the network $\mathsf{F}(\cdot,\cdot;\boldsymbol{\Theta}_{\mathsf{F}})$ dedicated to the fusion sub-task. The architecture of the network $\mathsf{C}(\cdot,\cdot;\boldsymbol{\Theta}_{\mathsf{C}})$ dedicated to CI inference only differs by the last ReLU layer that has been removed.}}
    \label{gene-change}
\end{figure}

This section provides details on the network architectures for a specific instance of the proposed framework. More precisely, it considers the particular case of complementary acquisitions, i.e., when the observed images $\mathbf{Y}_1$ and $\mathbf{Y}_2$ are LRHS and HRLS images respectively. 
As stated before, the fusion network $\mathsf{F}(\cdot,\cdot;\boldsymbol{\Theta}_{\mathsf{F}})$ can be chosen as any state-of-the-art architecture (or model-based algorithm)  \citep{PSGAN,wang2019deep,dian2020regularizing,xu2020hyperspectral,zhang2020deep,wang2021enhanced,hu2021hyperspectral} and is assumed to have been trained (or calibrated) beforehand. 
{In this work, as an example but without loss of generality, we adopt a network similar to the one introduced in \citep{PSGAN}, whose detailed architecture is depicted in Fig. \ref{gene-change}. 
Specifically, the LRHS and HRLS images are provided as inputs of two respective sub-networks designed for feature extraction.} 
Each sub-network consists of two successive convolutional layers followed by a leaky rectified linear unit (LeakyReLU) \citep{maas2013rectifier} and a down-convolution layer. 
Since the spatial resolution of the LRHS image $\mathbf{Y}_1$ is lower than the one of the target latent image $\mathbf{X}_1$, the second convolution layer of the dedicated sub-network is an up-convolution layer. 
The two resulting feature maps are then concatenated and pass through a series of convolutional layers, along with skip connections \citep{ronneberger2015u}. Finally, ReLU is applied in the last layer to ensure nonnegativity of the output fused image.

Regarding the network $\mathsf{C}(\cdot,\cdot;\boldsymbol{\Theta}_{\mathsf{C}})$ dedicated to CI inference, it basically performs a mapping with the same input and target spaces as the fusion network detailed above. 
Even if the respective underlying tasks are different (i.e., CD vs. fusion), their objectives are similar, e.g., extracting relevant spatial-spectral information from a pair of images with different spatial and spectral resolutions to produce a HRHS (latent or change) image. 
Thus, it is reasonable to adopt a similar architecture for $\mathsf{C}(\cdot,\cdot;\boldsymbol{\Theta}_{\mathsf{C}})$ as for $\mathsf{F}(\cdot,\cdot;\boldsymbol{\Theta}_{\mathsf{F}})$. 
The only difference lies in the last layer: the ReLU  has been removed, as  the CI is not necessarily nonnegative.

\begin{figure}[h!]
\centering
\includegraphics[width=0.5\linewidth]{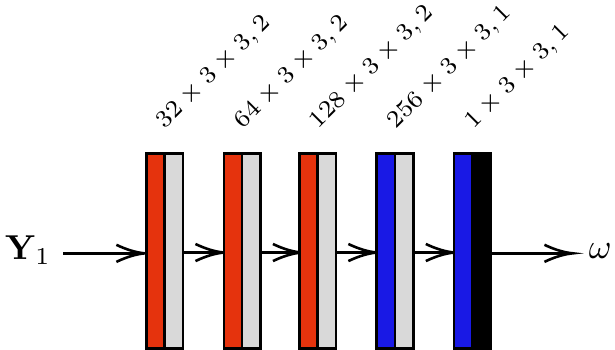}
    \caption{{Architecture of the discriminative network.}}
    \label{dis-net}
\end{figure}

Finally, the discriminative network $\mathsf{D}(\cdot;\boldsymbol{\Theta}_{\mathsf{D}})$ takes an LRHS image as an input, the observed image ${\mathbf{Y}}_1$ or the predicted image $\hat{\mathbf{Y}}_1$, to provide a binary decision $\omega \in \{0,1\}$. 
{When $\omega=1$, the network decides the input image is an actually observed one. Conversely, the discriminator provides the output $\omega=0$ when it decides the input image is a predicted image.} As depicted in Fig. \ref{dis-net}, it consists of three down-convolution layers and two-flat convolution layers. {The last layer of the network includes} a sigmoid activation function.

\section{Experiments on simulated data sets}\label{sec:experiments_synth}
{Statistical performance assessment of newly developed methods requires ground truth data to compute quantitative figures-of-merit. In the context of CD, ground truth data should provide the location of the changes that actually occurred in a geographical area between the two image acquisitions. However, in practice, it is generally difficult to obtain enough pairs of real images acquired
by heterogeneous sensors over the same geographical area at different times. Moreover, it is even more difficult to obtain the associated ground truth binary maps locating the changes between the acquisitions. As an alternative, this section aims at assessing the efficiency of the proposed CD-GAN framework thanks to experiments conducted on synthetic data. It compares its performance with standard and state-of-the-art CD techniques.  In this perspective, the proposed framework is instantiated for the particular case of complementary acquisitions. 
More precisely, throughout this section, the LRHS observed image $\mathbf{Y}_1$ is an HS image of low spatial resolution, while the HRLS observed image $\mathbf{Y}_2$ is an MS image of higher spatial resolution.}

\subsection{Synthetic data generation}
\label{subsec:data_generation}
{The experiments reported  hereafter have been conducted on synthetic data obtained following the simulation protocol proposed in \citep{ferraris2017detecting}.} 
This protocol relies on the availability of a single HRHS image $\mathbf{X}_{\textrm{ref}}$ to generate pairs of observed\footnote{To lighten the writting, we took the licence to refer to the synthetically generated HRLS and LRHS images as ``observed'' or ``acquired'' images.} HRLS and LRHS images through an unmixing-upmixing process. 
{This process introduces physically-inspired, realistic changes to the observed images, with predefined change maps that will be subsequently used to evaluate the performance of CD methods.} 
The successive steps of this protocol are briefly sketched below. 
Interested readers are invited to refer to the work in \citep{ferraris2017detecting} for more details.\\

\begin{figure}[!htp]
\centering
\includegraphics[width=0.15\textwidth]{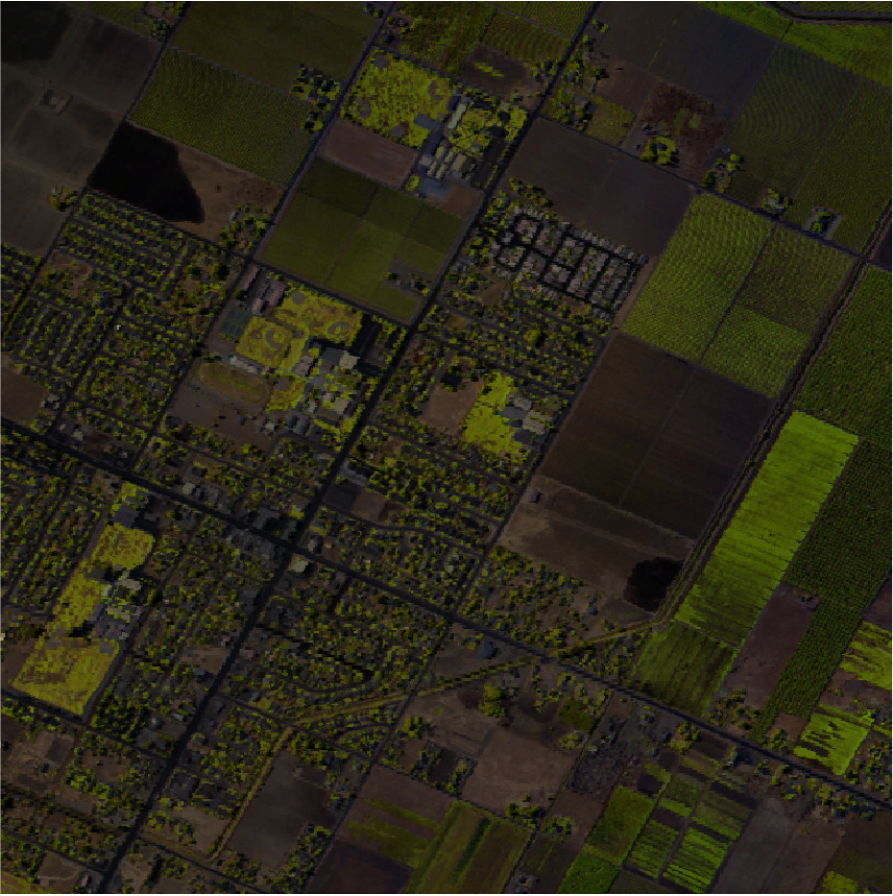}
\includegraphics[width=0.15\textwidth]{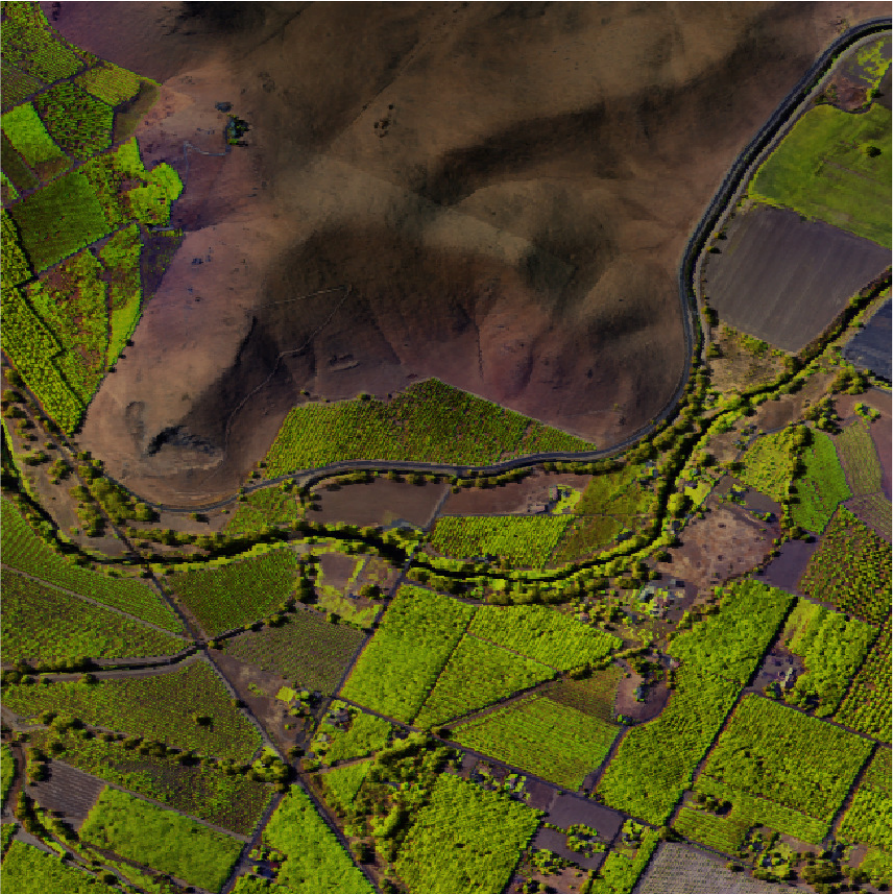}
\includegraphics[width=0.15\textwidth]{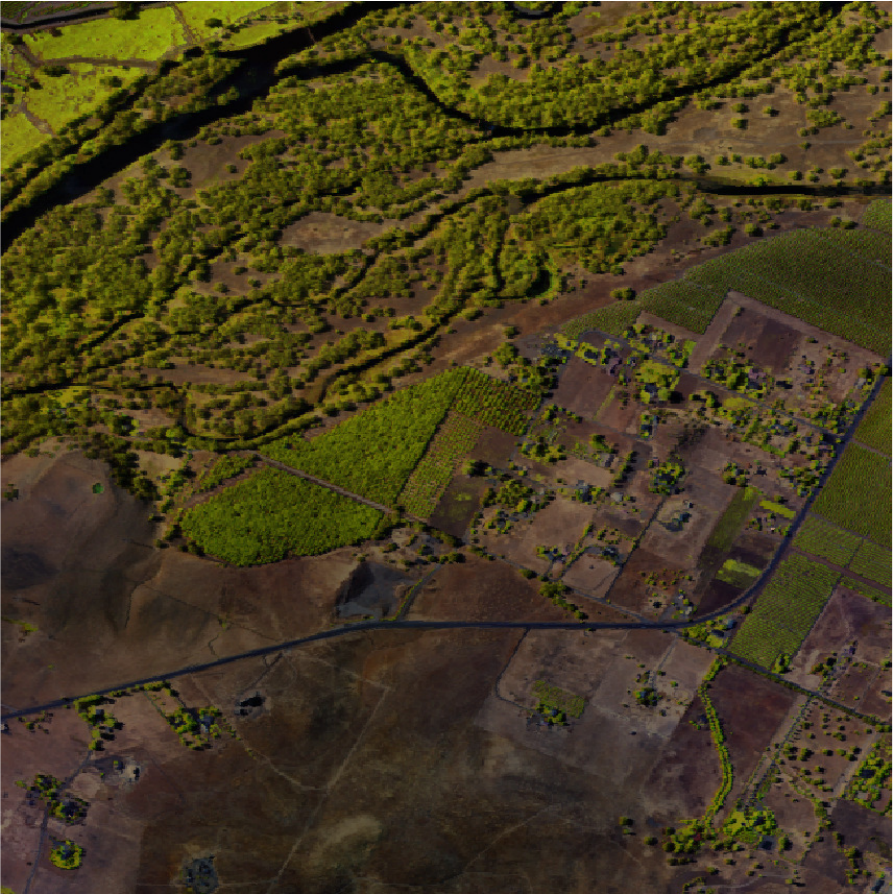}
\caption{Color composition of the three hyperspectral reference images.}
\label{highrs}
\end{figure}

\noindent \textbf{Generation of HRHS images --} Three sets of 22 HRHS reference images have been extracted from three real HS images acquired by the AVIRIS sensor, composed of $789\times 789$ pixels and $224$ spectral bands. They are depicted in Fig. \ref{highrs}. 
Each reference image denoted by $\mathbf{X}_{\textrm{ref}}$ is composed of $m=224$ spectral bands and are of size $120 \times 120$ pixels (i.e., $n=14400$). 
The resulting $66$ reference images are then processed as follows.\\

\noindent \textbf{Unmixing --} From each reference image $\mathbf{X}_{\textrm{ref}}$, a set of $k$ endmembers gathered in an $m \times k$ matrix $\mathbf{M}_{\textrm{ref}}$ are extracted using vertex component analysis \citep{VCA}, where the number $k$ of endmembers has been adjusted using Hysime \citep {Hysime}. Based on this set of endmembers, the reference image has been unmixed using SUnSAL \citep{SUnSAL} to recover a corresponding $k \times n$ abundance matrix $\mathbf{A}_{\textrm{ref}}$.\\

\noindent \textbf{Change generation --} Three kinds of simulated yet realistic changes are independently applied to the reference abundance matrices $\mathbf{A}_{\textrm{ref}}$ to produce modified abundance matrices denoted $\mathbf{A}_{\textrm{chg}}$. These three change rules are referred to as \emph{zero abundance} ($\texttt{R}_{\textrm{z}}$), \emph{same abundance} ($\texttt{R}_{\textrm{s}}$) and \emph{block abundance}  ($\texttt{R}_{\textrm{b}}$) in the sequel (see \citep{ferraris2017detecting} for more details). For each reference image, the pixels whose respective abundance vectors have been modified are identified by non-zero values in the predefined reference binary map $\mathbf{d}_{\textrm{ref}} \in \{0,1\}^n$.\\

\noindent \textbf{Upmixing --} Pairs $\left(\mathbf{X}_1,\mathbf{X}_2\right)$ of HRHS latent images are computed by linearly mixing endmember matrices $\mathbf{M}_{\textrm{ref}}$ with the abundances in $\mathbf{A}_{\textrm{ref}}$ or $\mathbf{A}_{\textrm{chg}}$. More precisely, these simulated pairs of latent images $\mathbf{X}_1$ and $\mathbf{X}_2$ are defined as
	\begin{equation}
	    \left(\mathbf{X}_1,\mathbf{X}_2\right) \in 
	    \left\{\left(\mathbf{M}_{\textrm{ref}}\mathbf{A}_{\textrm{ref}},\mathbf{M}_{\textrm{ref}}\mathbf{A}_{\textrm{chg}}\right),
	    \left(\mathbf{M}_{\textrm{ref}}\mathbf{A}_{\textrm{chg}},\mathbf{M}_{\textrm{ref}}\mathbf{A}_{\textrm{ref}}\right)\right\}.
	\end{equation}
 
\noindent \textbf{Generation of observed images --} Given the HRHS latent images $\mathbf{X}_1$ and $\mathbf{X}_2$ produced as above, respective pairs $\left(\mathbf{Y}_1,\mathbf{Y}_2\right)$ of observed images are generated according to the forward models \eqref{eq:directmodels}. In these experiments, as stated above, the observed images are assumed to be of complementary resolutions. Thus, the spatial degradation operator {$\mathcal{H}_1(\cdot)$} is chosen as a spatially-invariant Gaussian blur with standard deviation $\sigma=2.35$ followed  by a regular down-sampling in both directions of factor $4$ leading to $n_{2}= 16n_{1}$ (see \eqref{eq:dim}). The spectral degradation operator $\mathcal{H}_2(\cdot)$ mimics the response of a MS sensor by averaging four contiguous bands along the spectral dimension.\\

Finally, with the above steps, 396 pairs of images have been generated, from which 376 pairs have been used for training, and the other for testing.

\subsection{Experimental settings}
The proposed CD-GAN architecture is implemented using the PyTorch framework on a computer equipped with a Quadro RTX 6000 GPU. 
Unless otherwise stated, the hyperparameters adjusting the terms associated with the prediction and the spatial regularizations in the loss function \eqref{eq:overallloss} have been set to 
$\alpha=1$ and $\beta = 10^{-3}$. During the training, Adam \citep{2014Adam} is considered as the optimizer with an initial learning rate set to $2 \times 10^{-4}$ over 15 epochs and batches composed of $4$ training samples. These parameters have been adopted for all experiments reported in what follows. The change maps estimated by the proposed method is denoted as $\hat{\mathbf{d}}_{\textrm{CD-GAN}}$.\\

\noindent \textbf{Compared methods --} The proposed method is compared to five other ones. However, very few works considered the problem of CD between images of different resolutions. Most methods apply independent preprocessing to the two observed images to reach the same spatial and the same spectral resolutions which makes a pixel-wise comparison possible. 
\begin{itemize}
    \item The first method is the fusion-based approach proposed in \citep{ferraris2017detecting}. It first fuses the two observed images using the fast algorithm proposed in \citep{wei2015fast} to infer a common latent HRHS image. Then, an LRHS change image is computed from the spatially degraded latent image and the observed LRHS image ${\mathbf{Y}}_{1}$. Finally, a spatially regularized CVA (sCVA)  \citep{johnson1998change} has been conducted to produce an estimated LR CM denoted $\hat{\mathbf{d}}_{\textrm{F}}$.
    {\item The second method elaborates on the first one but it embeds the spatial–spectral reconstruction network for hyperspectral and multispectral image fusion (SSR-NET) \citep{SSRNET}. The estimated LR CM is denoted $\hat{\mathbf{d}}_{\textrm{F-SSRNET}}$.%
    \item The third method is the image translation-based unsupervised change detection proposed in \citep{Cycle-GAN}. The generated change map is denoted as $\hat{\mathbf{d}}_{\textrm{USCD}}$.}
    \item The fourth method consists in applying the superresolution algorithm proposed in \citep{zhao2016fast} to each band of the observed image $\mathbf{Y}_{1}$. The result is then spectrally degraded by applying $\mathcal{H}_2(\cdot)$. The resulting HRLS image can then be compared pixel-by-pixel to the observed image $\mathbf{Y}_2$ since they have the same spatial and the same spectral resolutions. The estimated HR change map $\hat{\mathbf{d}}_{\textrm{HRLS}}$ is thus obtained through spatially regularized change vector analysis (sCVA).
    \item  The fifth method applies the same operations as the second one but in a reverse order. The observed image $\mathbf{Y}_{1}$ is first spectrally degraded by applying $\mathcal{H}_2(\cdot)$ and then spatially superresolved using \citep{zhao2016fast}. The pair composed of the resulting HRLS image and the observed image $\mathbf{Y}_2$ is then analyzed using sCVA to produce a HR change map  denoted $\hat{\mathbf{d}}_{\textrm{LSHR}}$.
\end{itemize}

\noindent \textbf{Figures-of-merit --} To quantitatively evaluate the performance of the CD methods, the estimated change maps $\hat{\mathbf{d}}$ are compared to the reference change map $\mathbf{d}_{\textrm{ref}}$. 
This allows us to derive the empirical receiver operating characteristics (ROC) curves.
These ROC curves represent the estimated probability of detection $\mathrm{P}_{\textrm{D}}$ as a function of the estimated probability of false alarm $\mathrm{P}_{\textrm{FA}}$. 
Detection occurs when an actually changed pixel is identified as changed, while false alarm occurs when an unchanged pixel is identified as changed. 
Empirical ROC curves are the privileged figure-of-merit for detection performance assessment \citep{pepe2000receiver}. 
Contrary to classification-oriented metrics, a ROC curve comprehensively displays the estimated detection performance in terms of possible trade-offs between detection and false alarms on a single plot. 
Considering a wide range of threshold values allows to cover these possible trade-offs. 
Two quantitative metrics are also derived from these ROC curves, namely the area under the curve (AUC), sometimes referred to as the c-statistic \citep[Chap. 9]{Hastie2009}, and the distance (dist.) between the no detection point and the point at the interception of the ROC curve. 
For both metrics, the closer the values are to $1$, the better the CD methods.

\subsection{Results}
\begin{figure*}[h]
	\centering
	\includegraphics[width=0.32\textwidth]{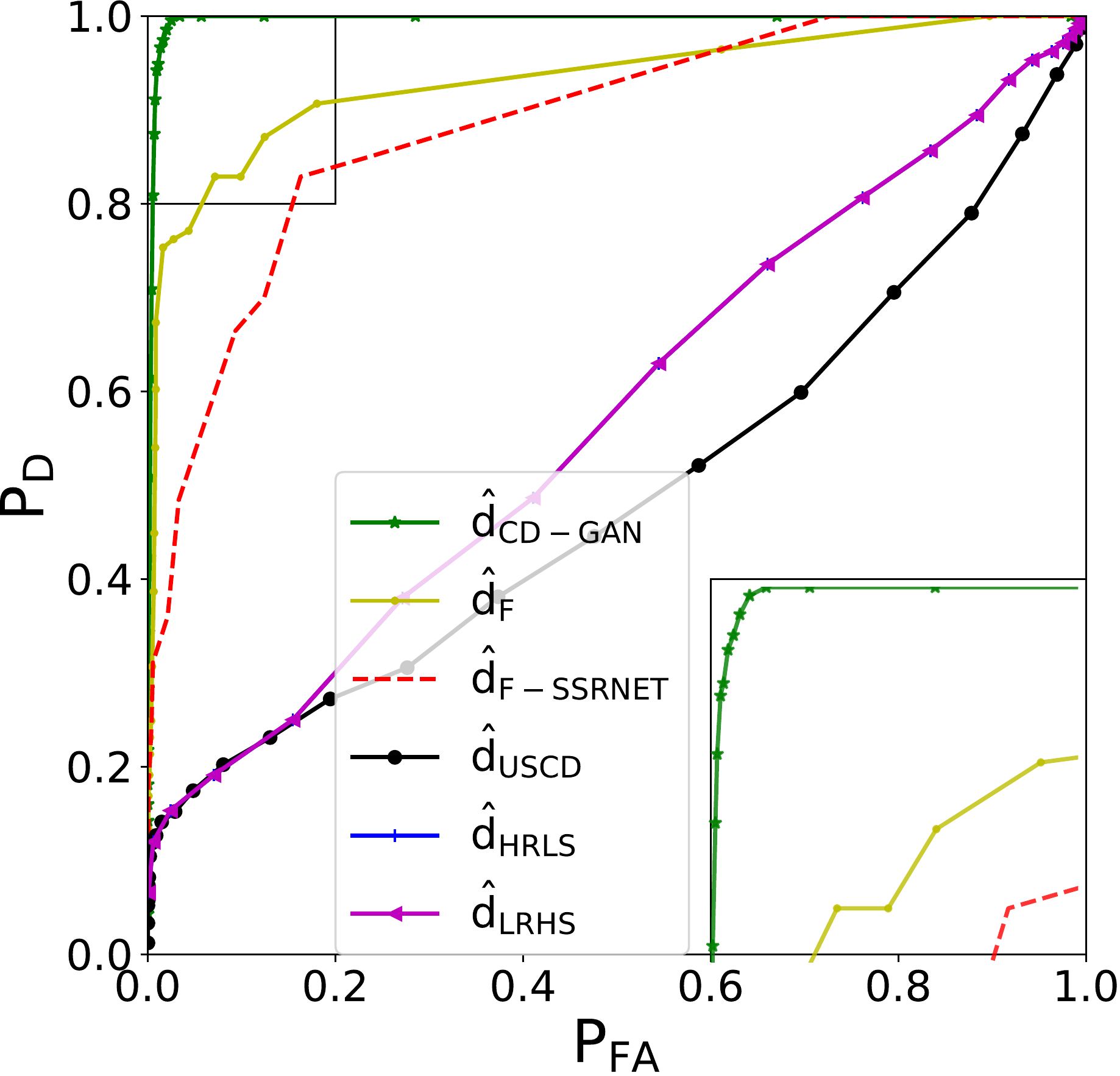}
	\includegraphics[width=0.32\textwidth]{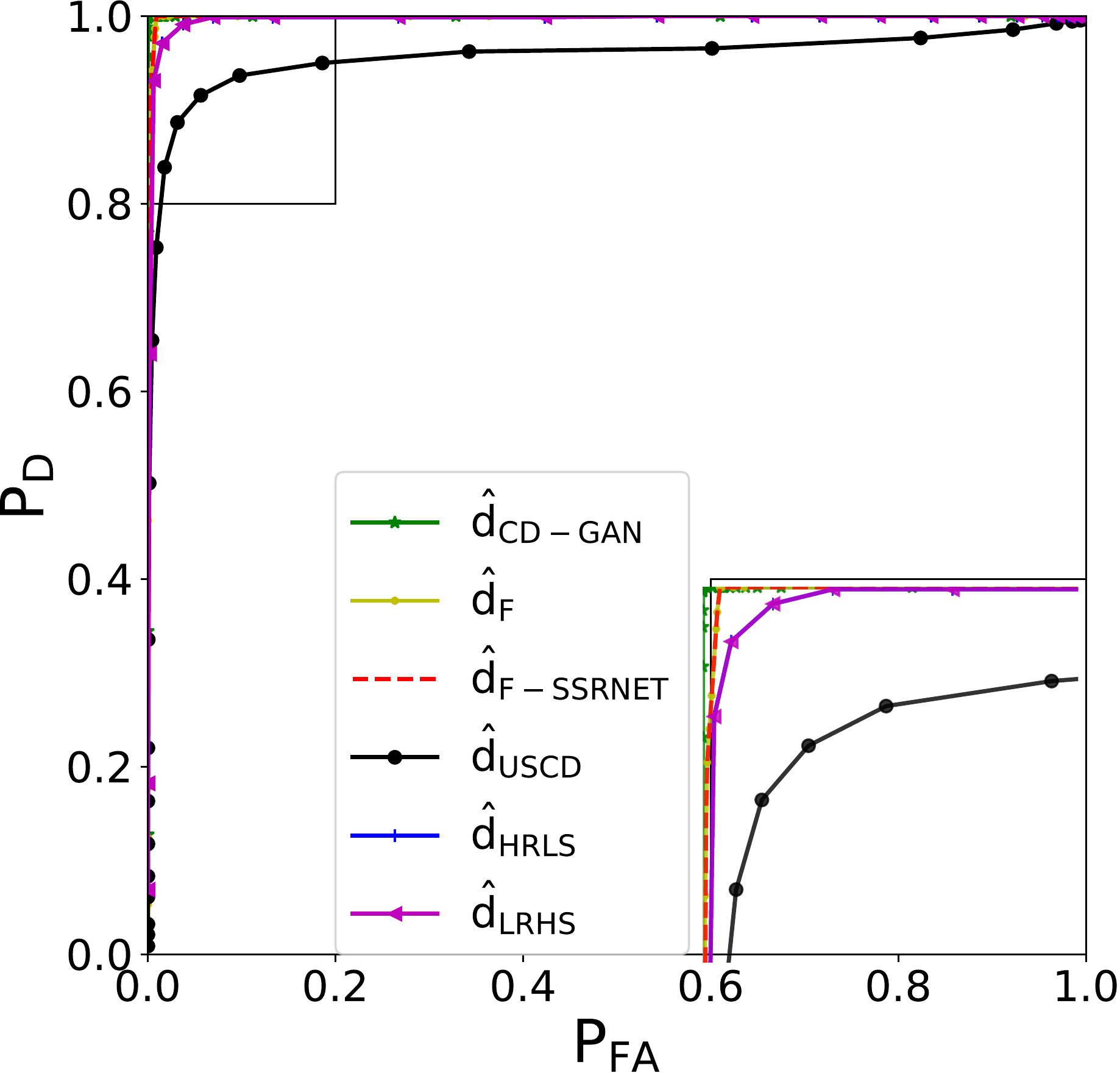}
	\includegraphics[width=0.32\textwidth]{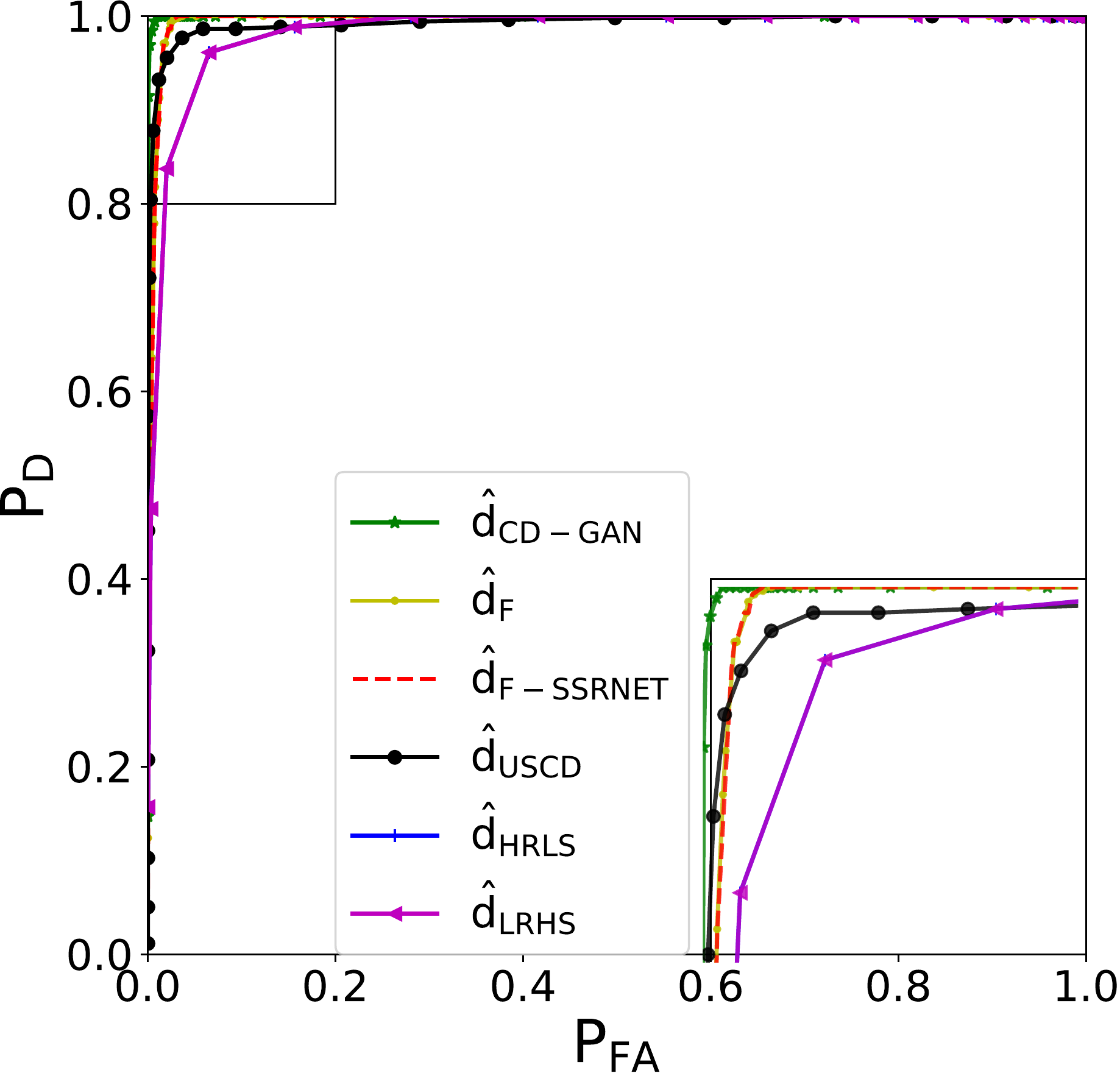}
	\caption{{ROC curves obtained by the compared CD methods for the three change rules: $\texttt{R}_b$ (left),  $\texttt{R}_s$ (middle) and  $\texttt{R}_z$ (right).}}
	\label{fake_ROC}
\end{figure*}

\begin{figure*}
	\centering
	\includegraphics[width=0.95\textwidth]{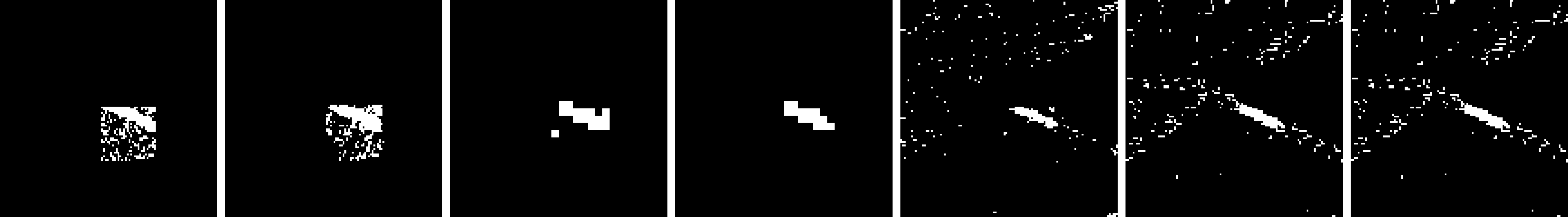}\\
	\includegraphics[width=0.95\textwidth]{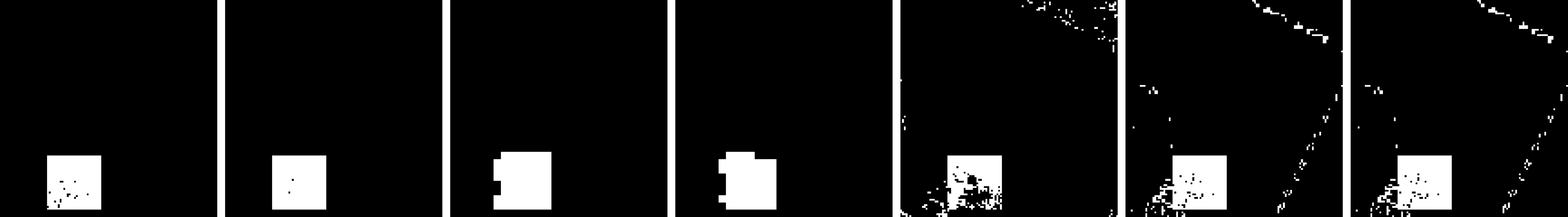}\\
	\includegraphics[width=0.95\textwidth]{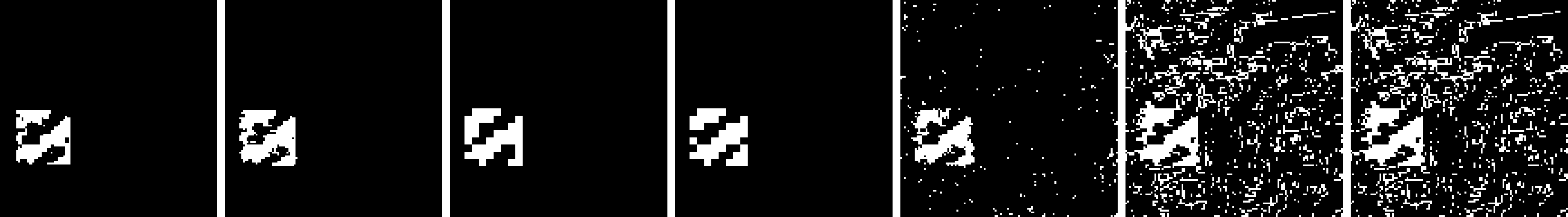}
	\caption{{CM estimated by the compared methods on the data sets generated according to the three change rules, namely $\texttt{R}_{\textrm{b}}$ (1st row), $\texttt{R}_{\textrm{s}}$ (2nd row) and $\texttt{R}_{\textrm{z}}$ (3rd row) with from left to right: ground truth, $\hat{\mathbf{d}}_{\textrm{CD-GAN}}$, $\hat{\mathbf{d}}_{\textrm{F}}$, $\hat{\mathbf{d}}_{\textrm{F-SSRNET}}$, $\hat{\mathbf{d}}_{\textrm{USCD}}$, $\hat{\mathbf{d}}_{\textrm{HRLS}}$, $\hat{\mathbf{d}}_{\textrm{LSHR}}$.}}
	\label{fake_CM}
\end{figure*}

\noindent \textbf{Performance comparison --} The ROC curves obtained by the compared methods are shown in Fig. \ref{fake_ROC}, for the three change rules $\texttt{R}_{\mathrm{b}}$, $\texttt{R}_{\mathrm{s}}$ and $\texttt{R}_{\mathrm{z}}$. 
The associated quantitative results are reported in Table \ref{fake-compare}. {Clearly, these first results show the superiority of the proposed CD-GAN framework when compared to the five other CD methods. 
More precisely, CD-GAN provides better detection even for very low probability of false alarm with an acceptable time. 
The methods based on $\hat{\mathbf{d}}_{\textrm{HRLS}}$ and on $\hat{\mathbf{d}}_{\textrm{LSHR}}$ are faster than other methods, behave similarly
and underperfom the other methods. }
Fig. \ref{fake_CM} shows the true CM and the CM estimated by the compared methods.  
Note that the estimated CMs  $\hat{\mathbf{d}}_{\textrm{F}}$  $\hat{\mathbf{d}}_{\textrm{F-SSRNET}}$ are defined at low spatial resolutions, contrary to the other ones.

\renewcommand\arraystretch{1.2}
\begin{table}[h!]
\centering
	
	\setlength{\tabcolsep}{0.8mm}{
		\begin{tabular}{c|c|c|c|c|c|c|c|}
			\cline{3-8} 
		\multicolumn{2}{c|}{}                                                              &$\hat{\mathbf{d}}_{\textrm{CD-GAN}}$       &$\hat{\mathbf{d}}_{\textrm{F}}$     &$\hat{\mathbf{d}}_{\textrm{F-SSRNET}}$  &$\hat{\mathbf{d}}_{\textrm{USCD}}$      &$\hat{\mathbf{d}}_{\textrm{HRLS}}$             &$\hat{\mathbf{d}}_{\textrm{LSHR}}$          \\
			\hline 	\hline 
			\multicolumn{1}{|c|}{\multirow{2}{*}{$\texttt{R}_{\textrm{b}}$}}      &AUC     &\textbf{0.9962}    &0.9660       &0.9342              &0.6906               &0.8081            &0.8081                 \\
\cline{2-8}  \multicolumn{1}{|c|}{}                                               &Dist.   &\textbf{0.9774}    &0.9190       &0.9324              &0.6906               &0.7248           &0.7248          \\
			\hline \hline
			\multicolumn{1}{|c|}{\multirow{2}{*}{$\texttt{R}_{\textrm{s}}$}}      &AUC     &\textbf{0.9999}    &0.9986       &0.9984              &0.9626               &0.9969             &0.9965                \\
\cline{2-8} \multicolumn{1}{|c|}{}                                                &Dist.   &\textbf{0.9988}    &0.9875      &0.9984              &0.9626               &0.9841             &0.9841             \\
	       \hline \hline	
			\multicolumn{1}{|c|}{\multirow{2}{*}{$\texttt{R}_{\textrm{z}}$}}      &AUC     &\textbf{0.9997}    &0.9956       &0.9957              &0.9930               &0.9844             &0.9843               \\
\cline{2-8} \multicolumn{1}{|c|}{}                                                &Dist.   &\textbf{0.9941}    &0.9864       &0.9957              &0.9930              &0.9612             &0.9612              \\	
			\hline \hline
			\multicolumn{2}{|c|}{Time(s)}      &1.9079    &0.3899       &0.2414              &0.1667               &0.0169             &0.0099               \\

\hline \hline			
			
	\end{tabular}}
		\caption{{Quantitative detection performance and computational times}.\label{fake-compare}}
\end{table}

\noindent \textbf{Sensitivity analysis --} We now discuss the impact of the three terms in the loss function in \eqref{eq:overallloss}, namely, the adversarial loss $\mathcal{L}_{\mathrm{adv}}\left(\cdot,\cdot\right)$, the prediction loss $ \mathcal{L}_{\mathrm{pre}}\left(\cdot\right)$ and the spatial sparsity regularization $\mathcal{L}_{\mathrm{spa}}\left(\cdot\right)$. Table \ref{beta-quan} reports the quantitative results associated with the three change rules $\texttt{R}_{\textrm{z}}$, $\texttt{R}_{\textrm{s}}$ and $\texttt{R}_{\textrm{b}}$ while adopting different loss terms in \eqref{eq:overallloss}: first column without the adversarial loss, second column without the prediction loss and third one without spatial sparsity regularization. Optimal values are highlighted in bold. Clearly, in this case, the prediction loss and the spatial sparsity regularization play important roles in the proposed CD-GAN, whereas combining the three terms leads to the best results.

In addition, we also analyze in this table the impact of the spatial sparsity regularization in the loss function \eqref{eq:overallloss}  by adjusting the hyperparameter $\beta$. The empirical results demonstrate the interest of incorporating this regularization since the case $\beta=0$, i.e., with no regularization, leads to the worst detection performance. Conversely, for a quite wide range of non-zero values of $\beta$, i.e., $\beta \in \{10^{-4}, 10^{-3}, 10^{-2} \}$, the detection performance is clearly better with the best one obtained for $\beta = 10^{-3}$, which is the default value chosen for the experiments reported in this section.

\renewcommand\arraystretch{1.2}
\begin{table}[h!]
\centering
	
	\setlength{\tabcolsep}{0.5mm}{
		\begin{tabular}{c|c|c|c|c|c|c|c|}
  			\cline{3-8}
  			\multicolumn{2}{c|}{} & no adv.        &$\alpha=0$      &$\beta=0$       &$\beta=10^{-4}$       &$\beta=10^{-3}$ & $\beta=10^{-2}$      \\
			\hline 	\hline 
			\multicolumn{1}{|c|}{\multirow{2}{*}{$\texttt{R}_{\textrm{b}}$}}      &AUC  &0.9954 &0.9324   &0.9335           &0.9925            &\textbf{0.9962}     &    0.9950        \\
\cline{2-8}  \multicolumn{1}{|c|}{} &Dist.  &0.9762 &0.9102  &0.8357           &    0.9688         &\textbf{0.9774}     &0.9651          \\
			\hline \hline
			\multicolumn{1}{|c|}{\multirow{2}{*}{$\texttt{R}_{\textrm{s}}$}}       &AUC   &0.9910 &0.9158   &0.8399           &     0.9835         &\textbf{0.9999}     &0.9896          \\
\cline{2-8} \multicolumn{1}{|c|}{}  &Dist.  &0.9718 &0.9031  &0.7892           &   0.9601            &\textbf{0.9999}     &0.9560       \\
			\hline \hline	
			\multicolumn{1}{|c|}{\multirow{2}{*}{$\texttt{R}_{\textrm{z}}$}}       &AUC  &9{0.9971} &0.9273    &0.8854           &       0.9768           &\textbf{0.9983}     &0.9937      \\
\cline{2-8}  \multicolumn{1}{|c|}{} &Dist. &0.9724 &0.9165   &0.8534           &   0.9350          &\textbf{0.9866}     &0.9684         \\	
			\hline \hline 
	\end{tabular}}
		\caption{Sensitivity analysis w.r.t. the hyperparameters in CD-GAN: quantitative detection performance.\label{beta-quan}}
\end{table}

Note that a similar analysis has been conducted for the hyperparameter $\alpha$ adjusting the weight of the prediction term in the overall loss \eqref{eq:overallloss}. The results, which show the limited impact of this hyperparameter on the detection performance, are not reported herein for brevity.\\

\noindent \textbf{Impact of the fusion methods --} In the proposed CD-GAN framework, the fusion network $\mathsf{F}\left(\cdot,\cdot;\boldsymbol{\Theta}_{\mathsf{F}}\right)$ can be chosen by the end-user and has been assumed to be  trained beforehand. Section \ref{subsec:archi_complementary_acquisitions} describes the fusion network adopted in most of the experiments discussed herein. This network heavily relies on the PS-GAN architecture proposed in \citep{PSGAN}. To illustrate its versatility, we propose to instantiate the proposed CD framework with two other fusion networks, namely the deep blind hyperspectral image fusion (DBIN) proposed in \citep{wang2019deep} and the SSR-NET network proposed in \citep{SSRNET}. Besides, as a complementary analysis, we consider a semi-supervised scenario where the latent image $\mathbf{X}_1$ (i.e., the HRHS image associated with the observed image $\mathbf{Y}_1$) is also available for training. Note that this more favorable scenario is the one considered in \citep{PSGAN} to train PS-GAN. In this case, the discriminative network $\mathsf{D}(\cdot;\boldsymbol{\Theta}_{\mathsf{D}})$ and the prediction loss \eqref{eq:preloss} can be easily adapted accordingly to distinguish the estimated fused image $\hat{\mathbf{X}}_1$ from the true latent image $\mathbf{X}_1$. The quantitative metrics are reported in Table \ref{fusion-quan} for the three implemented change rules. These results show that, whatever the adopted fusion network, the proposed unsupervised CD-GAN framework reaches detection performance comparable to the one obtained in a more favorable semi-supervised scenario. In addition, they show that the proposed CD framework is quite robust to the choice of the fusion method.\newline

\renewcommand\arraystretch{1.2}
  \begin{table}[h!]
  \centering
  	
  	\setlength{\tabcolsep}{1.2mm}{
  		\begin{tabular}{c|c|c|c|c|c|}
  			\cline{3-6}
  			\multicolumn{2}{c|}{}   & \multicolumn{3}{c|}{{Unsupervised} CD-GAN} & Semi-supervised \\
  			\cline{3-5}
  			\multicolumn{2}{c|}{} & DBIN        &SSR-NET       &PS-GAN       & CD-GAN    \\
  			\hline 	\hline 
  			\multicolumn{1}{|c|}{\multirow{2}{*}{$\texttt{R}_{\textrm{b}}$}}      &AUC        &0.9860        &0.9962           &0.9962             &\textbf{0.9971}               \\
  	\cline{2-6} 
  	\multicolumn{1}{|c|}{}&Dist.  &0.9650    &0.9683      &\textbf{0.9774}             &0.9774             \\
  			\hline \hline
  			\multicolumn{1}{|c|}{\multirow{2}{*}{$\texttt{R}_{\textrm{s}}$}}       &AUC        &0.9933     &\textbf{0.9974}     &0.9952             &0.9946               \\
  	\cline{2-6} 
  	\multicolumn{1}{|c|}{}&Dist.  &\textbf{0.9750}     &0.9675      &0.9734             &0.9737             \\
  			\hline \hline	
  			\multicolumn{1}{|c|}{\multirow{2}{*}{$\texttt{R}_{\textrm{z}}$}}       &AUC        &0.9975    &\textbf{0.9984}       &0.9949             &0.9940              \\
    \cline{2-6} 
    \multicolumn{1}{|c|}{}&Dist.  &0.9849     &\textbf{0.9954}      &0.9758             &0.9851          \\	
  			\hline \hline
  	\end{tabular}}
  	  	\caption{Impact of the fusion method: quantitative detection performance.\label{fusion-quan}}
  \end{table}

\noindent \textbf{Robustness w.r.t. the forward models --} The proposed framework sketched in Fig. \ref{flowchart} requires the definition of the spatial and spectral degradation operators $\mathcal{H}_1(\cdot)$ and $\mathcal{H}_2(\cdot)$. In some practical applications, one may have only a partial knowledge regarding these operators related to the sensors. Experiments have been conducted to assess the robustness of the proposed CD-GAN framework w.r.t. model mismatch. More precisely, first the CD-GAN architecture has been trained with corrupted degradation operators on a training set composed of observed images generated with uncorrupted degradation operators and following the protocol described in Section \ref{subsec:data_generation}. Then, with all network parameters fixed, the corrupted operators have been updated in a second training procedure. Finally, the CD-GAN has been trained a second time based on these corrected operators. Regarding the corruption processes, the spatial degradation operator $\mathcal{H}_1(\cdot)$ specified in Section \ref{subsec:data_generation} has been replaced by a spatially-invariant Gaussian blur with standard deviation $\sigma= 1.70$. Inspired by the study conducted in \citep{Wei_IEEE_JSTSP_2015}, the spectral filters which compose the spectral degradation operator $\mathcal{H}_2(\cdot)$ have been corrupted by an additive zero-mean Gaussian noise with a variance leading to a $\textrm{SNR}=8$dB. 

\renewcommand\arraystretch{1.2}
\begin{table}[h!]
		\centering
	
	\setlength{\tabcolsep}{0.8mm}{
		\begin{tabular}{c|c|c|c|c|c|c|c|}
			\cline{3-8} 
		\multicolumn{2}{c|}{}                                                     &$\hat{\mathbf{d}}_{\textrm{CD-GAN}}$             &$\hat{\mathbf{d}}_{\textrm{F}}$    &$\hat{\mathbf{d}}_{\textrm{F-SSRNET}}$  &$\hat{\mathbf{d}}_{\textrm{USCD}}$       &$\hat{\mathbf{d}}_{\textrm{HRLS}}$             &$\hat{\mathbf{d}}_{\textrm{LSHR}}$  \\
			\hline 	\hline 
			\multicolumn{1}{|c|}{\multirow{2}{*}{$\texttt{R}_{\textrm{b}}$}}      &AUC     &\textbf{0.9948}    &0.7236       &0.8638         &0.4655            &0.5619      &0.5619           \\
\cline{2-8}  \multicolumn{1}{|c|}{}                                               &Dist.   &\textbf{0.9714}   &0.6423       &0.8638         &0.4655            &0.4862      &0.4862       \\
			\hline \hline
			\multicolumn{1}{|c|}{\multirow{2}{*}{$\texttt{R}_{\textrm{s}}$}}      &AUC     &\textbf{0.9918}    &0.9825       &0.9847         &0.9421            &0.9802     &0.9796           \\
\cline{2-8} \multicolumn{1}{|c|}{}                                                &Dist.   &\textbf{0.9876}    &0.9762       &0.9734         &0.9349             &0.9563     &0.9557         \\
	       \hline \hline	
			\multicolumn{1}{|c|}{\multirow{2}{*}{$\texttt{R}_{\textrm{z}}$}}      &AUC     &\textbf{0.9934}    &0.9753       &0.9811         &0.8017                  &0.7543     &0.7498           \\
\cline{2-8} \multicolumn{1}{|c|}{}                                                &Dist.   &\textbf{0.9812}    &0.9216       &0.9653         &0.017                  &0.6498     &0.6420         \\	
			\hline \hline 
	\end{tabular}}
		\caption{Robustness w.r.t. forward models: quantitative detection performance.\label{fake-quan}}
\end{table}

Table \ref{fake-quan} reports the quantitative results obtained by the compared methods. They show that the proposed CD-GAN obtains the highest metrics. In conclusion, when compared with other methods, the proposed CD-GAN framework is shown to be robust w.r.t. the misspecification of the degradation operators for the three implemented change rules. 

\section{Experiments on real data sets}\label{sec:experiments_real}
\begin{figure}[h!]
	\centering
	\includegraphics[width=0.15\textwidth]{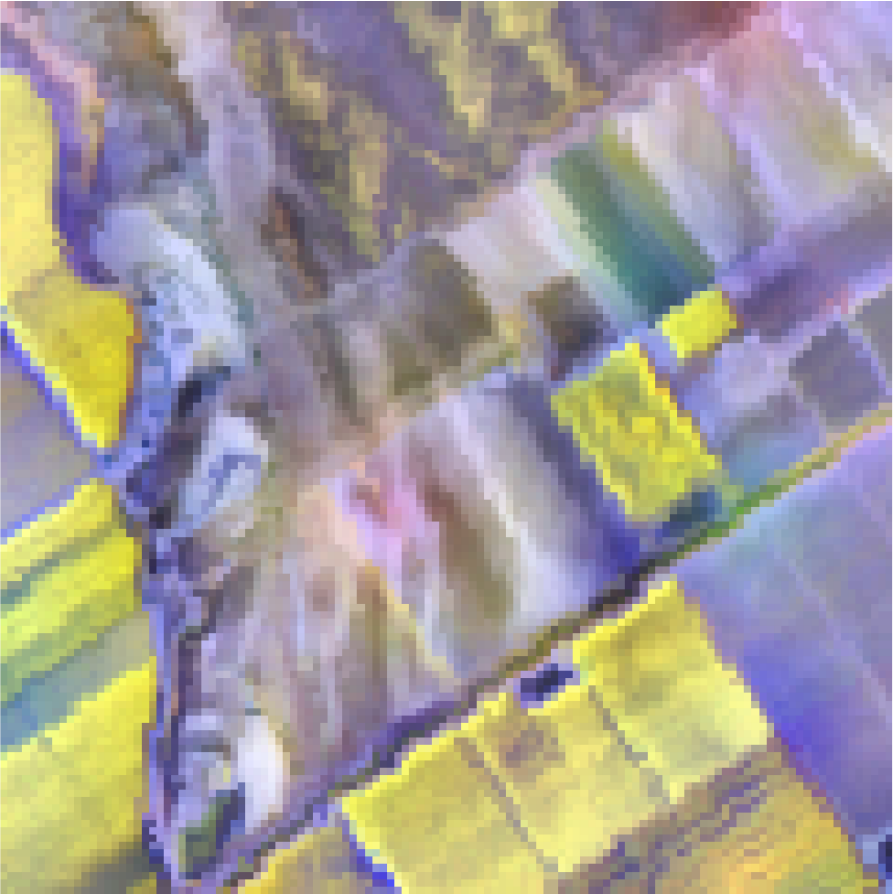}
	\includegraphics[width=0.15\textwidth]{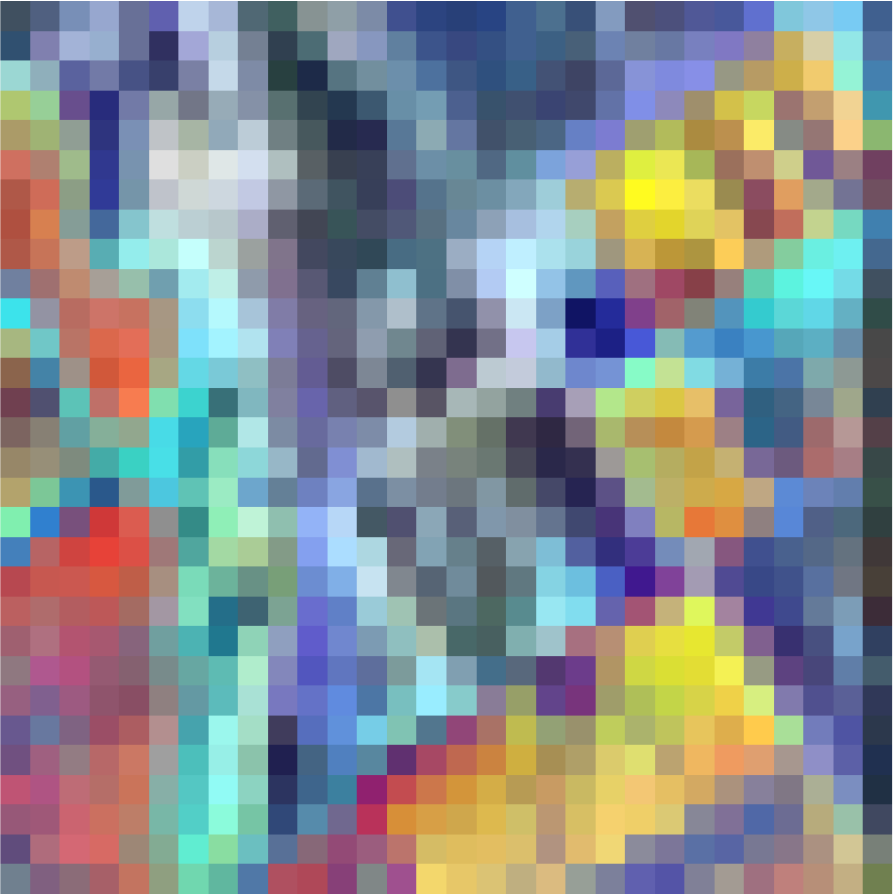}
	\includegraphics[width=0.15\textwidth]{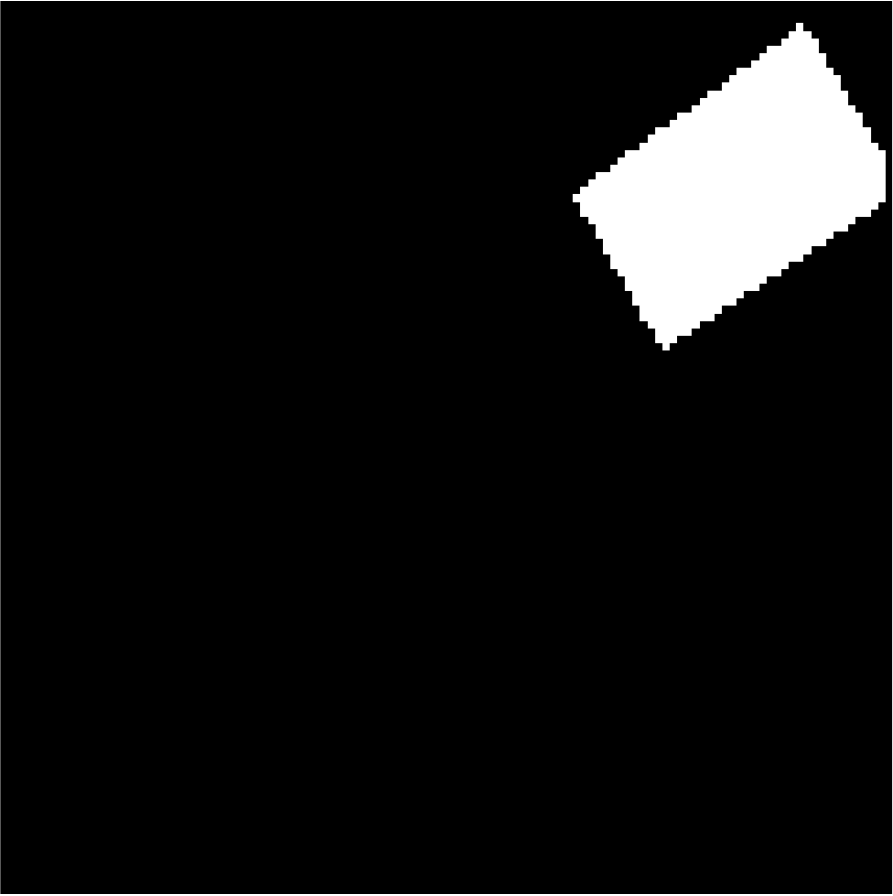}
	\\\vspace{0.1cm}
	\includegraphics[width=0.15\textwidth]{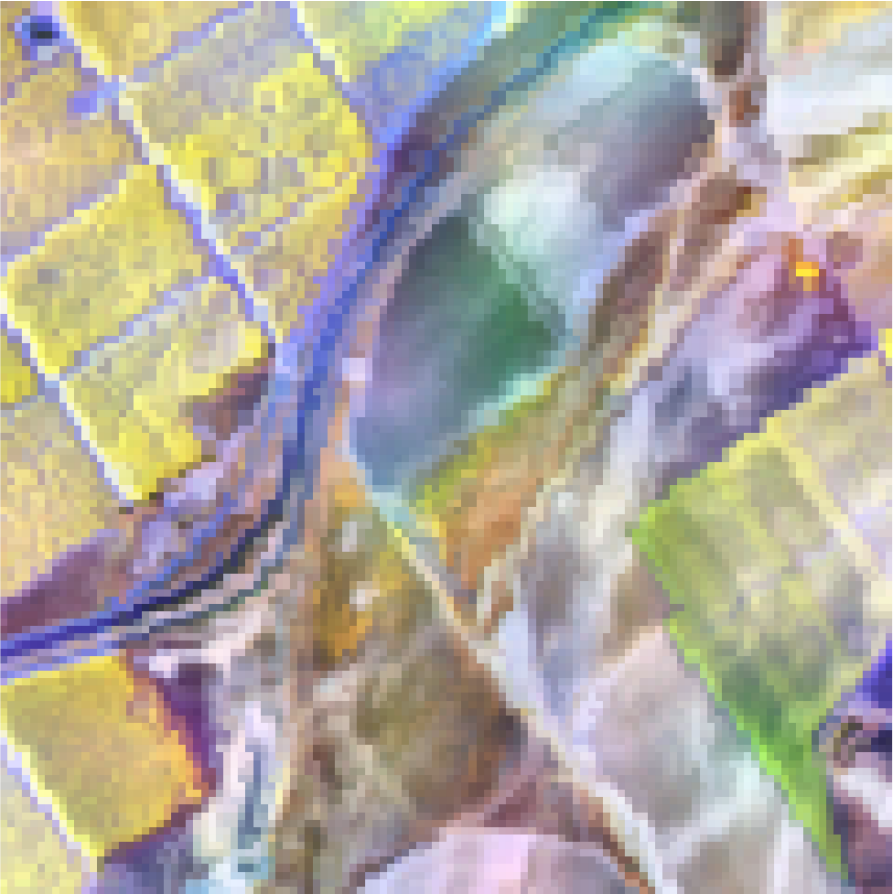}
	\includegraphics[width=0.15\textwidth]{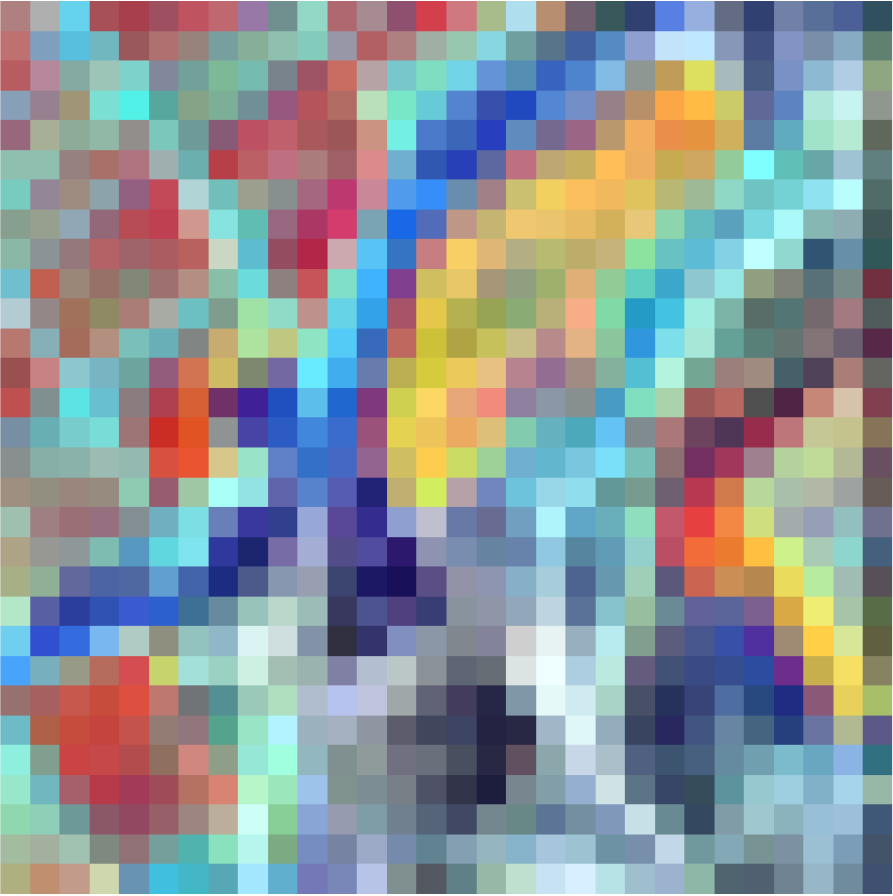}
	\includegraphics[width=0.15\textwidth]{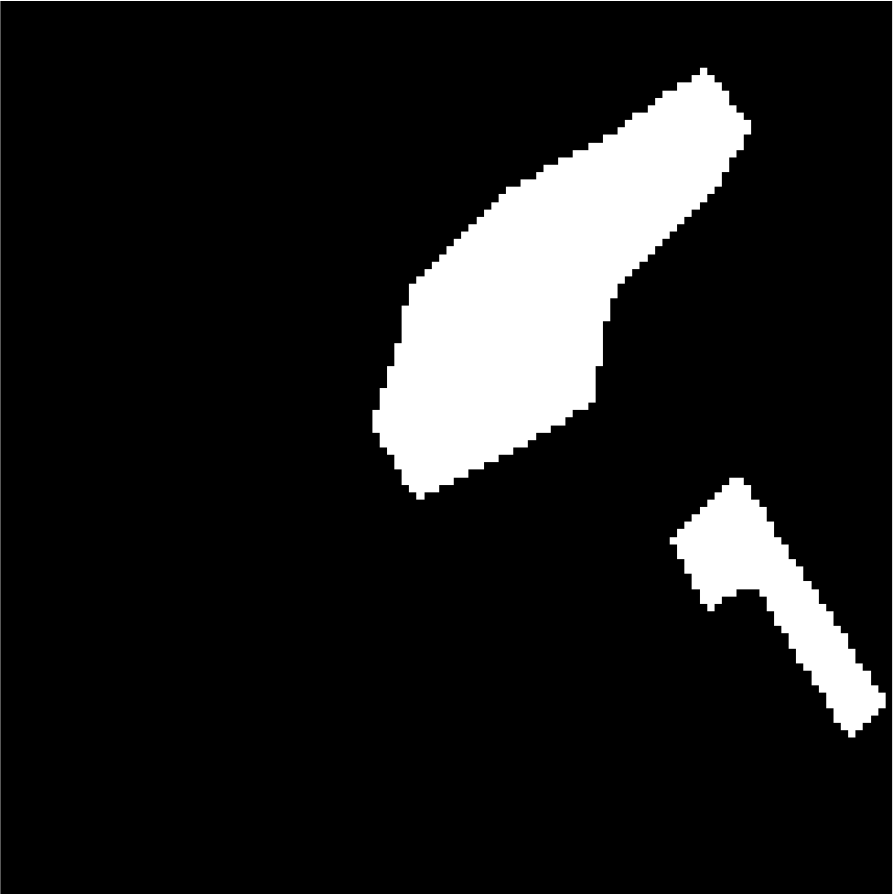}
\caption{Data sets \texttt{SB1} (top) and \texttt{SB2} (bottom): color composition of the HRLS observed image (left), color composition of the LRHS observed image (middle) and the binary CM (right) where white (resp. black) pixels correspond to changed (resp. unchanged) areas.}
\label{ref}
\end{figure}

For an illustrative purpose, complementary experiments have been conducted on a real data set, namely the Santa Barbara data set. 
It is composed of two HS images acquired by the AVIRIS sensor in 2013 and 2014, respectively, to monitor the change of composition over Santa Barbara area in California \citep{shafique2022deep}. 
These images are composed of $984 \times 740$ pixels, with $m=224$ spectral bands. In the experiments, two HRHS sub-images of sizes $120\times 120$ are selected. The spatial and spectral degradation operators $\mathcal{H}_1(\cdot)$ and $\mathcal{H}_2(\cdot)$ detailed in Section \ref{subsec:data_generation} are applied to each of these HRHS images to mimic heterogeneous acquisitions with complementary resolutions, i.e., producing one LRHS image and one HRLS image. The HRLS and LRHS images as well as the actual binary CMs constituting the two data sets, denoted \texttt{SB1} and \texttt{SB2}, are shown in Fig. \ref{ref}. The binary CMs are designed as in \citep{shafique2022deep}.

\begin{figure}[h!]
\centering
	\includegraphics[width=0.49\textwidth]{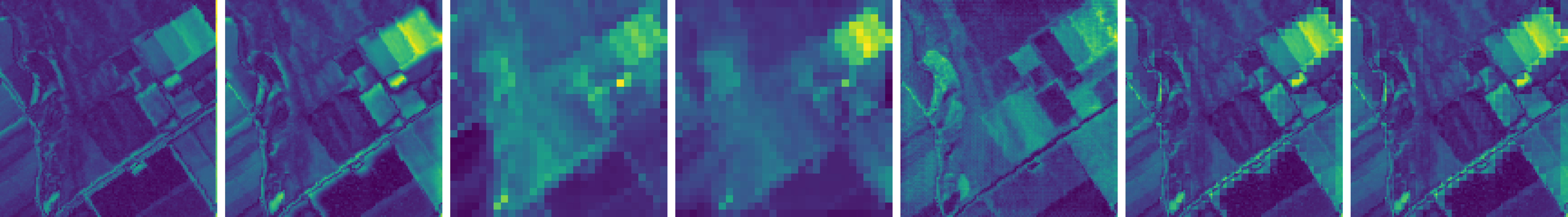}\\
	\includegraphics[width=0.49\textwidth]{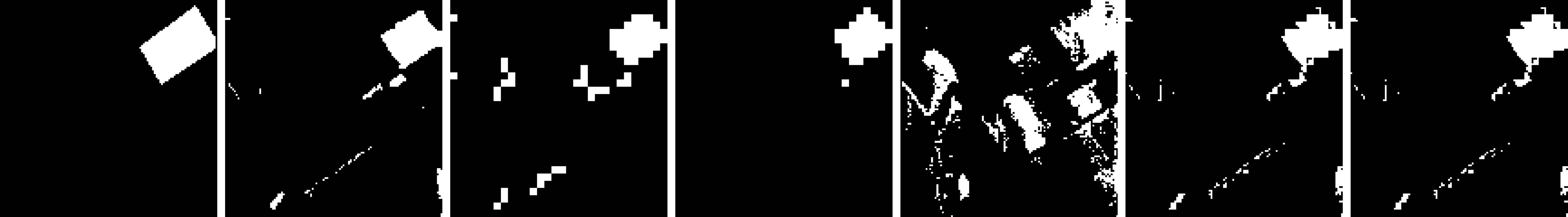}
	\caption{{Data set \texttt{SB1}: CIs (top) and CM (bottom) estimated by the compared methods with, from left to right: ground truth, $\hat{\mathbf{d}}_{\textrm{CD-GAN}}$, $\hat{\mathbf{d}}_{\textrm{F}}$, $\hat{\mathbf{d}}_{\textrm{F-SSRNET}}$, $\hat{\mathbf{d}}_{\textrm{USCD}}$, $\hat{\mathbf{d}}_{\textrm{HRLS}}$, $\hat{\mathbf{d}}_{\textrm{LSHR}}$.}}
	\label{fig:SB1}
\end{figure}

\begin{figure}[h!]
	\centering
	\includegraphics[width=0.225\textwidth]{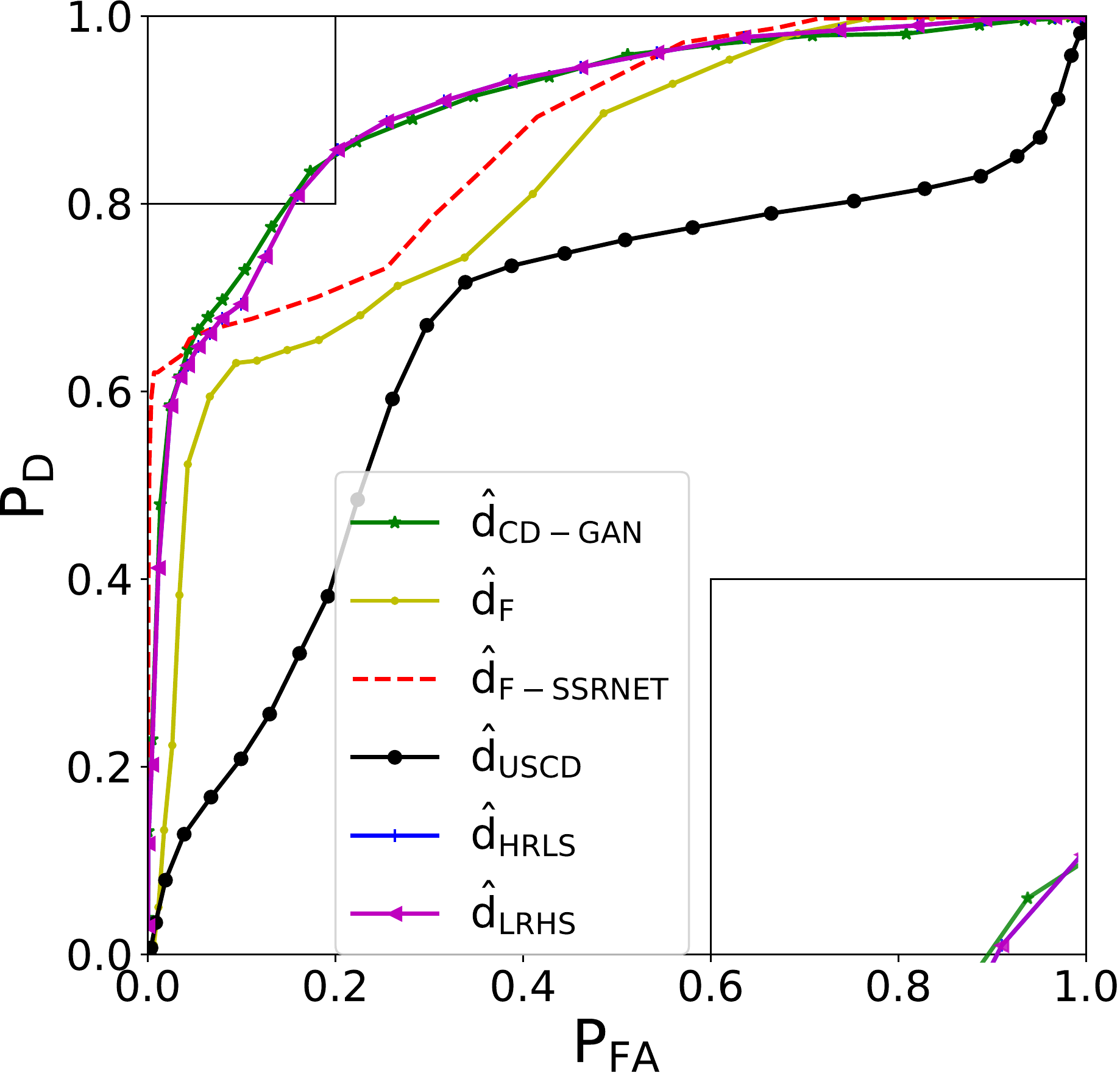}
	\includegraphics[width=0.225\textwidth]{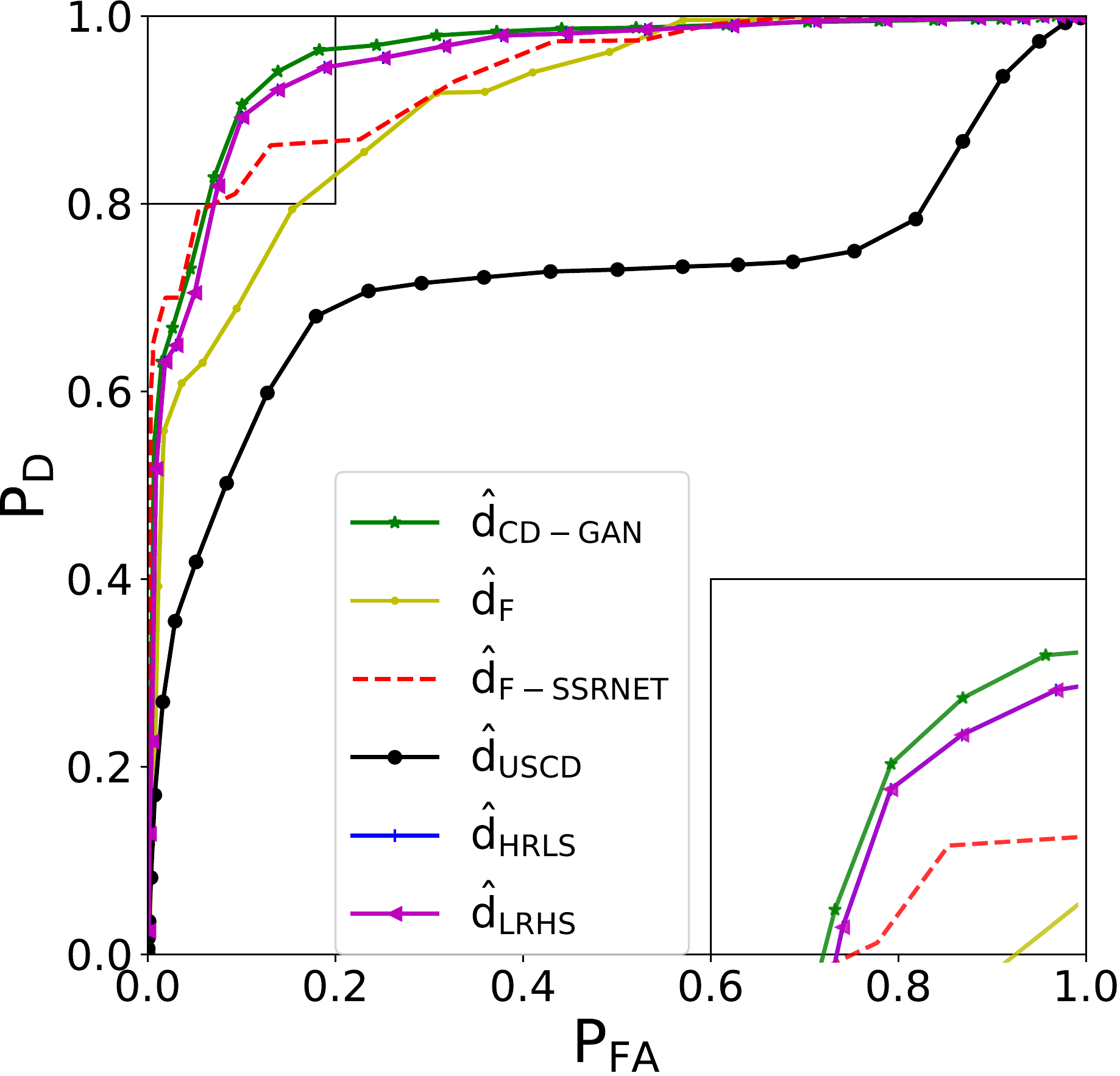}
	\caption{Data sets \texttt{SB1} (left) and \texttt{SB2} (right): ROC curves obtained by the compared methods.}
	\label{fig:SB1_ROC}
\end{figure}

Regarding the data set \texttt{SB1}, the CIs and the resulting binary CMs recovered by the compared methods are shown in Fig. \ref{fig:SB1}. 
Visual inspection allows one to state that recovering the changed areas between the two observed images is a challenging task for all compared methods. The estimated CM $\hat{\mathbf{d}}_{\textrm{USCD}}$ contains a lot of false alarms, i.e., pixels decided as changed which actually are not affected by any change. The other compared methods exhibit significantly less false alarms, in particular, the estimated CM $\hat{\mathbf{d}}_{\textrm{F-SSRNET}}$. However it is worth noting that the change maps $\hat{\mathbf{d}}_{\textrm{F}}$ and $\hat{\mathbf{d}}_{\textrm{F-SSRNET}}$ are defined at low spatial resolution. The results associated with the proposed CD-GAN show accurate detection of the changes with a low number of false alarms. These findings are confirmed by the ROC curves depicted in Fig. \ref{fig:SB1_ROC} (left) and the quantitative results reported in Table \ref{tab:SB1_quan}. From Fig. \ref{fig:SB1_ROC} (left), it appears that the proposed CD-GAN provides high detection rate whatever the functioning point of the detector (i.e., for all values of PFA). From a quantitative point-of-view, the proposed CD-GAN framework obtains higher dist. score and AUC.\newline

\renewcommand\arraystretch{1.2}
\begin{table}[h!]
	\centering
	\label{tab:SB1_quan}
	\setlength{\tabcolsep}{0.6mm}{
		\begin{tabular}{c|c|c|c|c|c|c|c|}
	\cline{3-8}
	\multicolumn{2}{c|}{}                                                     &$\hat{\mathbf{d}}_{\textrm{CD-GAN}}$        &$\hat{\mathbf{d}}_{\textrm{F}}$   &$\hat{\mathbf{d}}_{\textrm{F-SSRNET}}$  &$\hat{\mathbf{d}}_{\textrm{USCD}}$   &$\hat{\mathbf{d}}_{\textrm{HRLS}}$          &$\hat{\mathbf{d}}_{\textrm{LSHR}}$        \\
\hline 	\hline 
\multicolumn{1}{|c|}{\multirow{2}{*}{\texttt{SB1}}}  &AUC                     &\textbf{0.9035}                             &0.8263                             &0.8753                                      &0.6519                                     &0.9033                                &0.9033         \\
\cline{2-8} \multicolumn{1}{|c|}{}                   &Dist.                   &\textbf{0.8342}                             &0.7124                            &0.8753                                      &0.6519                                     &0.8091                                &0.8091             \\
\hline \hline
\multicolumn{1}{|c|}{\multirow{2}{*}{\texttt{SB2}}} &AUC                      &\textbf{0.9577}                             &0.9075                            & 0.9370                                     &0.7199                                     &0.9496                                &0.9498                \\
\cline{2-8} \multicolumn{1}{|c|}{}                  &Dist.                    &\textbf{0.9057}                             &0.7939                            & 0.9370                                     &0.7199                                     &0.8923                                &0.8923             \\	
\hline \hline
	\end{tabular}}
		\caption{{Data sets \texttt{SB1} and \texttt{SB2}: quantitative detection performance.}}
\end{table}

\begin{figure}[h!]
	\centering
 \includegraphics[width=0.49\textwidth]{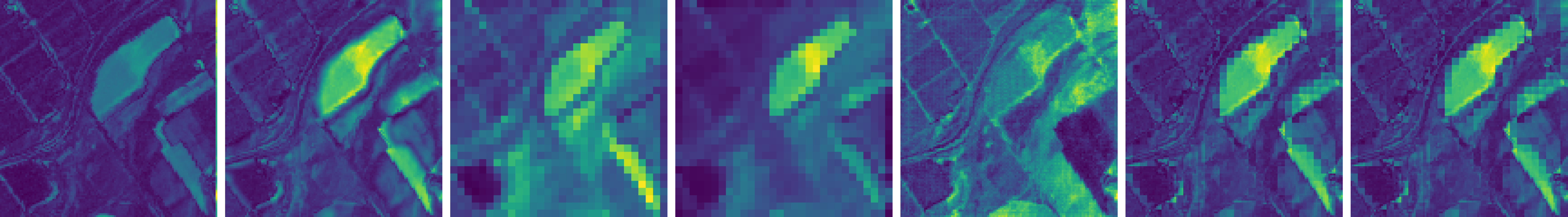}\\
      \includegraphics[width=0.49\textwidth]{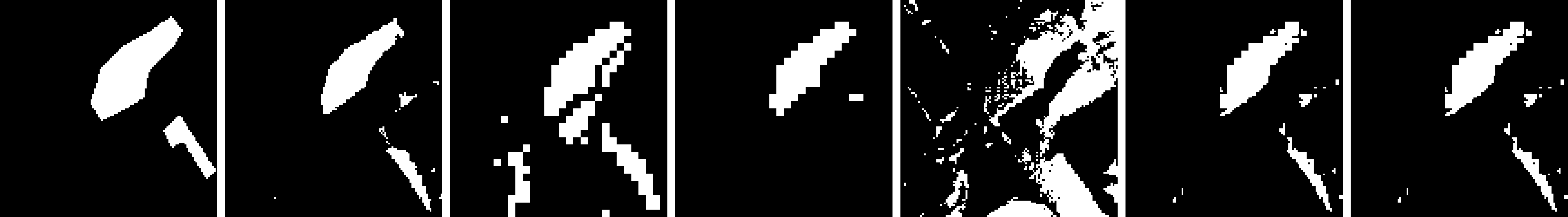}\\     
	\caption{{Data set \texttt{SB2}: CIs (top) and CM (bottom)  estimated by the compared methods with, from left to right: ground truth, $\hat{\mathbf{d}}_{\textrm{CD-GAN}}$, $\hat{\mathbf{d}}_{\textrm{F}}$, $\hat{\mathbf{d}}_{\textrm{F-SSRNET}}$, $\hat{\mathbf{d}}_{\textrm{USCD}}$,$\hat{\mathbf{d}}_{\textrm{HRLS}}$, $\hat{\mathbf{d}}_{\textrm{LSHR}}$.}}
\label{Yancheng-data}
\end{figure}

Regarding the data set \texttt{SB2}, the CI and CM estimated by the compared methods are shown in Fig. \ref{Yancheng-data}. For the middle part, the change maps $\hat{\mathbf{d}}_{\textrm{F}}$ and $\hat{\mathbf{d}}_{\textrm{F-USCD}}$ are composed of many isolated pixels wrongly detected as changed.  The estimated  CM $\hat{\mathbf{d}}_{\textrm{F-SSRNET}}$ fails to detect large part of changed pixels. The CM estimated by the proposed CD-GAN framework is composed of  more well-located changed pixels, with a significant reduction of false alarms. This visual inspection is confirmed by the quantitative results reported in Table \ref{tab:SB1_quan} where the proposed CD-GAN framework reaches higher metrics (AUC and dist.) than the compared methods. The ROC curves depicted in Fig. \ref{fig:SB1_ROC} (right) show that the CD-GAN strategy provides better detection whatever the functioning point of the detector, i.e., for a wide range of false alarm probability.

\section{Conclusion}\label{sec:conclusion}
This paper introduced a robust fusion-based adversarial framework for detecting changes between heterogeneous images in an unsupervised scenario.
The proposed approach capitalized on the availability of a predefined and previously trained fusion network. Following a robust fusion-based change detection strategy, this network was complemented with another network having the same architecture and specifically dedicated to detect change between images of possibly different spatial and spectral resolutions. The overall architecture was trained within an adversarial paradigm, enriching the canonical adversarial loss function with task-driven terms.
Experiments conducted on simulated and real data sets illustrated the efficiency, the versatility and the robustness of the proposed change detection framework.

\section*{Acknowledgements}
The authors would like to thank Florentin Coeurdoux (IRIT, University of Toulouse) for his relevant advice regarding the practical implementation of the proposed CD-GAN architecture.

\bibliographystyle{elsarticle-harv}

\bibliography{strings_all_ref,ref}

\begin{thebibliography}{58}
\expandafter\ifx\csname natexlab\endcsname\relax\def\natexlab#1{#1}\fi
\providecommand{\url}[1]{\texttt{#1}}
\providecommand{\href}[2]{#2}
\providecommand{\path}[1]{#1}
\providecommand{\DOIprefix}{doi:}
\providecommand{\ArXivprefix}{arXiv:}
\providecommand{\URLprefix}{URL: }
\providecommand{\Pubmedprefix}{pmid:}
\providecommand{\doi}[1]{\href{http://dx.doi.org/#1}{\path{#1}}}
\providecommand{\Pubmed}[1]{\href{pmid:#1}{\path{#1}}}
\providecommand{\bibinfo}[2]{#2}
\ifx\xfnm\relax \def\xfnm[#1]{\unskip,\space#1}\fi
\bibitem[{Alberga et~al.(2007a)Alberga, Idrissa, Lacroix and
  Inglada}]{alberga2007comparison}
\bibinfo{author}{Alberga, V.}, \bibinfo{author}{Idrissa, M.},
  \bibinfo{author}{Lacroix, V.}, \bibinfo{author}{Inglada, J.},
  \bibinfo{year}{2007}a.
\newblock \bibinfo{title}{Comparison of similarity measures of multi-sensor
  images for change detection applications}, in: \bibinfo{booktitle}{Proc. IEEE
  Int. Conf. Geosci. Remote Sens. (IGARSS)}, \bibinfo{organization}{IEEE}. pp.
  \bibinfo{pages}{2358--2361}.
\bibitem[{Alberga et~al.(2007b)Alberga, Idrissa, Lacroix and
  Inglada}]{alberga2007performance}
\bibinfo{author}{Alberga, V.}, \bibinfo{author}{Idrissa, M.},
  \bibinfo{author}{Lacroix, V.}, \bibinfo{author}{Inglada, J.},
  \bibinfo{year}{2007}b.
\newblock \bibinfo{title}{Performance estimation of similarity measures of
  multi-sensor images for change detection applications}, in:
  \bibinfo{booktitle}{Proc. IEEE Int. Workshop Analysis Multitemporal Remote
  Sensing Images (MultiTemp)}, \bibinfo{organization}{IEEE}. pp.
  \bibinfo{pages}{1--5}.
\bibitem[{Bovolo and Bruzzone(2006)}]{bovolo2006theoretical}
\bibinfo{author}{Bovolo, F.}, \bibinfo{author}{Bruzzone, L.},
  \bibinfo{year}{2006}.
\newblock \bibinfo{title}{A theoretical framework for unsupervised change
  detection based on change vector analysis in the polar domain}.
\newblock \bibinfo{journal}{IEEE Trans. Geosci. Remote Sens.}
  \bibinfo{volume}{45}, \bibinfo{pages}{218--236}.
\bibitem[{Bruzzone and Bovolo(2012)}]{bruzzone2012novel}
\bibinfo{author}{Bruzzone, L.}, \bibinfo{author}{Bovolo, F.},
  \bibinfo{year}{2012}.
\newblock \bibinfo{title}{A novel framework for the design of change-detection
  systems for very-high-resolution remote sensing images}.
\newblock \bibinfo{journal}{Proc. IEEE} \bibinfo{volume}{101},
  \bibinfo{pages}{609--630}.
\bibitem[{Bruzzone and Prieto(2002)}]{bruzzone2002adaptive}
\bibinfo{author}{Bruzzone, L.}, \bibinfo{author}{Prieto, D.F.},
  \bibinfo{year}{2002}.
\newblock \bibinfo{title}{An adaptive semiparametric and context-based approach
  to unsupervised change detection in multitemporal remote-sensing images}.
\newblock \bibinfo{journal}{IEEE Trans. Image Process.} \bibinfo{volume}{11},
  \bibinfo{pages}{452--466}.
\bibitem[{Dian et~al.(2023)Dian, Guo and Li}]{dian2023zero}
\bibinfo{author}{Dian, R.}, \bibinfo{author}{Guo, A.}, \bibinfo{author}{Li,
  S.}, \bibinfo{year}{2023}.
\newblock \bibinfo{title}{Zero-shot hyperspectral sharpening}.
\newblock \bibinfo{journal}{IEEE Trans. Patt. Anal. Mach. Intell.} .
\bibitem[{Dian et~al.(2020)Dian, Li and Kang}]{dian2020regularizing}
\bibinfo{author}{Dian, R.}, \bibinfo{author}{Li, S.}, \bibinfo{author}{Kang,
  X.}, \bibinfo{year}{2020}.
\newblock \bibinfo{title}{Regularizing hyperspectral and multispectral image
  fusion by {CNN} denoiser}.
\newblock \bibinfo{journal}{IEEE Trans. Neural Netw. Learn. Syst.}
  \bibinfo{volume}{32}, \bibinfo{pages}{1124--1135}.
\bibitem[{Du et~al.(2019)Du, Ru, Wu and Zhang}]{du2019unsupervised}
\bibinfo{author}{Du, B.}, \bibinfo{author}{Ru, L.}, \bibinfo{author}{Wu, C.},
  \bibinfo{author}{Zhang, L.}, \bibinfo{year}{2019}.
\newblock \bibinfo{title}{Unsupervised deep slow feature analysis for change
  detection in multi-temporal remote sensing images}.
\newblock \bibinfo{journal}{IEEE Trans. Geosci. Remote Sens.}
  \bibinfo{volume}{57}, \bibinfo{pages}{9976--9992}.
\bibitem[{Du et~al.(2020)Du, Wang, Chen, Liu, Lin and Meng}]{du2020improved}
\bibinfo{author}{Du, P.}, \bibinfo{author}{Wang, X.}, \bibinfo{author}{Chen,
  D.}, \bibinfo{author}{Liu, S.}, \bibinfo{author}{Lin, C.},
  \bibinfo{author}{Meng, Y.}, \bibinfo{year}{2020}.
\newblock \bibinfo{title}{An improved change detection approach using
  tri-temporal logic-verified change vector analysis}.
\newblock \bibinfo{journal}{ISPRS J. Photogramm. Remote Sens.}
  \bibinfo{volume}{161}, \bibinfo{pages}{278--293}.
\bibitem[{Ferraris et~al.(2020)Ferraris, Dobigeon and
  Chabert}]{Ferraris_INFFUS_2020}
\bibinfo{author}{Ferraris, V.}, \bibinfo{author}{Dobigeon, N.},
  \bibinfo{author}{Chabert, M.}, \bibinfo{year}{2020}.
\newblock \bibinfo{title}{Robust fusion algorithms for unsupervised change
  detection between multi-band optical images -- {A} comprehensive case study}.
\newblock \bibinfo{journal}{Information Fusion} \bibinfo{volume}{64},
  \bibinfo{pages}{293--317}.
\bibitem[{Ferraris et~al.(2017a)Ferraris, Dobigeon, Wei and
  Chabert}]{ferraris2017detecting}
\bibinfo{author}{Ferraris, V.}, \bibinfo{author}{Dobigeon, N.},
  \bibinfo{author}{Wei, Q.}, \bibinfo{author}{Chabert, M.},
  \bibinfo{year}{2017}a.
\newblock \bibinfo{title}{Detecting changes between optical images of different
  spatial and spectral resolutions: a fusion-based approach}.
\newblock \bibinfo{journal}{IEEE Trans. Geosci. Remote Sens.}
  \bibinfo{volume}{56}, \bibinfo{pages}{1566--1578}.
\bibitem[{Ferraris et~al.(2017b)Ferraris, Dobigeon, Wei and
  Chabert}]{Ferraris_IEEE_Trans_CI_2017}
\bibinfo{author}{Ferraris, V.}, \bibinfo{author}{Dobigeon, N.},
  \bibinfo{author}{Wei, Q.}, \bibinfo{author}{Chabert, M.},
  \bibinfo{year}{2017}b.
\newblock \bibinfo{title}{Robust fusion of multi-band images with different
  spatial and spectral resolutions for change detection}.
\newblock \bibinfo{journal}{IEEE Trans. Comput. Imag.} \bibinfo{volume}{3},
  \bibinfo{pages}{175--186}.
\bibitem[{Gao et~al.(2016)Gao, Dong, Li and Xu}]{gao2016automatic}
\bibinfo{author}{Gao, F.}, \bibinfo{author}{Dong, J.}, \bibinfo{author}{Li,
  B.}, \bibinfo{author}{Xu, Q.}, \bibinfo{year}{2016}.
\newblock \bibinfo{title}{Automatic change detection in synthetic aperture
  radar images based on {PCANet}}.
\newblock \bibinfo{journal}{IEEE Geosci. Remote Sens. Lett.}
  \bibinfo{volume}{13}, \bibinfo{pages}{1792--1796}.
\bibitem[{Gueguen and Hamid(2016)}]{gueguen2016toward}
\bibinfo{author}{Gueguen, L.}, \bibinfo{author}{Hamid, R.},
  \bibinfo{year}{2016}.
\newblock \bibinfo{title}{Toward a generalizable image representation for
  large-scale change detection: {A}pplication to generic damage analysis}.
\newblock \bibinfo{journal}{IEEE Trans. Geosci. Remote Sens.}
  \bibinfo{volume}{54}, \bibinfo{pages}{3378--3387}.
\bibitem[{Hastie et~al.(2009)Hastie, Tibshirani and Friedman}]{Hastie2009}
\bibinfo{author}{Hastie, T.}, \bibinfo{author}{Tibshirani, R.},
  \bibinfo{author}{Friedman, J.H.}, \bibinfo{year}{2009}.
\newblock \bibinfo{title}{The elements of statistical learning: data mining,
  inference, and prediction}.
\newblock \bibinfo{edition}{2} ed., \bibinfo{publisher}{Springer},
  \bibinfo{address}{New York}.
\bibitem[{Hegazy and Kaloop(2015)}]{hegazy2015monitoring}
\bibinfo{author}{Hegazy, I.R.}, \bibinfo{author}{Kaloop, M.R.},
  \bibinfo{year}{2015}.
\newblock \bibinfo{title}{Monitoring urban growth and land use change detection
  with {GIS} and remote sensing techniques in {D}aqahlia governorate {E}gypt}.
\newblock \bibinfo{journal}{Int. J. Sustainable Built Environment}
  \bibinfo{volume}{4}, \bibinfo{pages}{117--124}.
\bibitem[{Hu et~al.(2021)Hu, Huang, Deng, Jiang, Vivone and
  Chanussot}]{hu2021hyperspectral}
\bibinfo{author}{Hu, J.F.}, \bibinfo{author}{Huang, T.Z.},
  \bibinfo{author}{Deng, L.J.}, \bibinfo{author}{Jiang, T.X.},
  \bibinfo{author}{Vivone, G.}, \bibinfo{author}{Chanussot, J.},
  \bibinfo{year}{2021}.
\newblock \bibinfo{title}{Hyperspectral image super-resolution via deep
  spatiospectral attention convolutional neural networks}.
\newblock \bibinfo{journal}{IEEE Trans. Neural Netw. Learn. Syst.} .
\bibitem[{Iordache et~al.(2011)Iordache, Bioucas-Dias and Plaza}]{SUnSAL}
\bibinfo{author}{Iordache, M.D.}, \bibinfo{author}{Bioucas-Dias, J.M.},
  \bibinfo{author}{Plaza, A.}, \bibinfo{year}{2011}.
\newblock \bibinfo{title}{Sparse unmixing of hyperspectral data}.
\newblock \bibinfo{journal}{IEEE Trans. Geosci. Remote Sens.}
  \bibinfo{volume}{49}, \bibinfo{pages}{2014--2039}.
\bibitem[{Isola et~al.(2017)Isola, Zhu, Zhou and Efros}]{isola2017image}
\bibinfo{author}{Isola, P.}, \bibinfo{author}{Zhu, J.Y.},
  \bibinfo{author}{Zhou, T.}, \bibinfo{author}{Efros, A.A.},
  \bibinfo{year}{2017}.
\newblock \bibinfo{title}{Image-to-image translation with conditional
  adversarial networks}, in: \bibinfo{booktitle}{Proc. Int. Conf. on Computer
  Vision and Pattern Recognition (CVPR)}, pp. \bibinfo{pages}{1125--1134}.
\bibitem[{Johnson and Kasischke(1998)}]{johnson1998change}
\bibinfo{author}{Johnson, R.D.}, \bibinfo{author}{Kasischke, E.},
  \bibinfo{year}{1998}.
\newblock \bibinfo{title}{Change vector analysis: A technique for the
  multispectral monitoring of land cover and condition}.
\newblock \bibinfo{journal}{Int. J. Remote Sensing} \bibinfo{volume}{19},
  \bibinfo{pages}{411--426}.
\bibitem[{Kingma and Ba(2014)}]{2014Adam}
\bibinfo{author}{Kingma, D.}, \bibinfo{author}{Ba, J.}, \bibinfo{year}{2014}.
\newblock \bibinfo{title}{Adam: A method for stochastic optimization}.
\newblock \bibinfo{journal}{Proc. IEEE Int. Conf. Learn. Represent. (ICLR)} .
\bibitem[{Li et~al.(2021)Li, Gong, Zhang and Wu}]{li2021spatially}
\bibinfo{author}{Li, H.}, \bibinfo{author}{Gong, M.}, \bibinfo{author}{Zhang,
  M.}, \bibinfo{author}{Wu, Y.}, \bibinfo{year}{2021}.
\newblock \bibinfo{title}{Spatially self-paced convolutional networks for
  change detection in heterogeneous images}.
\newblock \bibinfo{journal}{IEEE J. Sel. Topics Appl. Earth Observations Remote
  Sens.} \bibinfo{volume}{14}, \bibinfo{pages}{4966--4979}.
\bibitem[{Li et~al.(2023)Li, Dian and Liu}]{li2023learning}
\bibinfo{author}{Li, S.}, \bibinfo{author}{Dian, R.}, \bibinfo{author}{Liu,
  H.}, \bibinfo{year}{2023}.
\newblock \bibinfo{title}{Learning the external and internal priors for
  multispectral and hyperspectral image fusion}.
\newblock \bibinfo{journal}{Science China Information Sciences}
  \bibinfo{volume}{66}, \bibinfo{pages}{140303}.
\bibitem[{Li et~al.(2017)Li, Shi, Zhang and Hao}]{li2017change}
\bibinfo{author}{Li, Z.}, \bibinfo{author}{Shi, W.}, \bibinfo{author}{Zhang,
  H.}, \bibinfo{author}{Hao, M.}, \bibinfo{year}{2017}.
\newblock \bibinfo{title}{Change detection based on gabor wavelet features for
  very high resolution remote sensing images}.
\newblock \bibinfo{journal}{IEEE Geosci. Remote Sens. Lett.}
  \bibinfo{volume}{14}, \bibinfo{pages}{783--787}.
\bibitem[{Liu et~al.(2019)Liu, Gong, Qin and Tan}]{liu2019bipartite}
\bibinfo{author}{Liu, J.}, \bibinfo{author}{Gong, M.}, \bibinfo{author}{Qin,
  A.K.}, \bibinfo{author}{Tan, K.C.}, \bibinfo{year}{2019}.
\newblock \bibinfo{title}{Bipartite differential neural network for
  unsupervised image change detection}.
\newblock \bibinfo{journal}{IEEE Trans. Neural Netw. Learn. Syst.}
  \bibinfo{volume}{31}, \bibinfo{pages}{876--890}.
\bibitem[{Liu et~al.(2016)Liu, Gong, Qin and Zhang}]{liu2016deep}
\bibinfo{author}{Liu, J.}, \bibinfo{author}{Gong, M.}, \bibinfo{author}{Qin,
  K.}, \bibinfo{author}{Zhang, P.}, \bibinfo{year}{2016}.
\newblock \bibinfo{title}{A deep convolutional coupling network for change
  detection based on heterogeneous optical and radar images}.
\newblock \bibinfo{journal}{IEEE Trans. Neural Netw. Learn. Syst.}
  \bibinfo{volume}{29}, \bibinfo{pages}{545--559}.
\bibitem[{Liu et~al.(2020)Liu, Zhou, Xu, Liu and Wang}]{PSGAN}
\bibinfo{author}{Liu, Q.}, \bibinfo{author}{Zhou, H.}, \bibinfo{author}{Xu,
  Q.}, \bibinfo{author}{Liu, X.}, \bibinfo{author}{Wang, Y.},
  \bibinfo{year}{2020}.
\newblock \bibinfo{title}{Psgan: A generative adversarial network for remote
  sensing image pan-sharpening}.
\newblock \bibinfo{journal}{IEEE Trans. Geosci. Remote Sens.} .
\bibitem[{Liu et~al.(2021)Liu, Zhang, Pan and Ning}]{Cycle-GAN}
\bibinfo{author}{Liu, Z.G.}, \bibinfo{author}{Zhang, Z.W.},
  \bibinfo{author}{Pan, Q.}, \bibinfo{author}{Ning, L.B.},
  \bibinfo{year}{2021}.
\newblock \bibinfo{title}{Unsupervised change detection from heterogeneous data
  based on image translation}.
\newblock \bibinfo{journal}{IEEE Trans. Geosci. Remote Sens.} ,
  \bibinfo{pages}{1--13}.
\bibitem[{Loncan et~al.(2015)Loncan, Almeida, {Bioucas-Dias}, Briottet,
  Chanussot, Dobigeon, Fabre, Liao, Licciardi, Simoes, Tourneret, Veganzones,
  Vivone, Wei and Yokoya}]{Loncan_IEEE_GRS_Mag_2015}
\bibinfo{author}{Loncan, L.}, \bibinfo{author}{Almeida, L.B.},
  \bibinfo{author}{{Bioucas-Dias}, J.M.}, \bibinfo{author}{Briottet, X.},
  \bibinfo{author}{Chanussot, J.}, \bibinfo{author}{Dobigeon, N.},
  \bibinfo{author}{Fabre, S.}, \bibinfo{author}{Liao, W.},
  \bibinfo{author}{Licciardi, G.}, \bibinfo{author}{Simoes, M.},
  \bibinfo{author}{Tourneret, J.Y.}, \bibinfo{author}{Veganzones, M.},
  \bibinfo{author}{Vivone, G.}, \bibinfo{author}{Wei, Q.},
  \bibinfo{author}{Yokoya, N.}, \bibinfo{year}{2015}.
\newblock \bibinfo{title}{Hyperspectral pansharpening: a review}.
\newblock \bibinfo{journal}{IEEE Geosci. Remote Sens. Mag.}
  \bibinfo{volume}{3}, \bibinfo{pages}{27--46}.
\bibitem[{Maas et~al.(2013)Maas, Hannun, Ng et~al.}]{maas2013rectifier}
\bibinfo{author}{Maas, A.L.}, \bibinfo{author}{Hannun, A.Y.},
  \bibinfo{author}{Ng, A.Y.}, et~al., \bibinfo{year}{2013}.
\newblock \bibinfo{title}{Rectifier nonlinearities improve neural network
  acoustic models}, in: \bibinfo{booktitle}{Proc. Int. Conf. Machine Learning
  (ICML)}, \bibinfo{organization}{Citeseer}. p.~\bibinfo{pages}{3}.
\bibitem[{Manonmani and Suganya(2010)}]{manonmani2010remote}
\bibinfo{author}{Manonmani, R.}, \bibinfo{author}{Suganya, G.},
  \bibinfo{year}{2010}.
\newblock \bibinfo{title}{Remote sensing and {GIS} application in change
  detection study in urban zone using multi temporal satellite}.
\newblock \bibinfo{journal}{Int. J. Geomatics and Geosci.} \bibinfo{volume}{1},
  \bibinfo{pages}{60--65}.
\bibitem[{Nascimento and Bioucas-Dias(2007)}]{Hysime}
\bibinfo{author}{Nascimento, J.}, \bibinfo{author}{Bioucas-Dias, J.M.},
  \bibinfo{year}{2007}.
\newblock \bibinfo{title}{Hyperspectral signal subspace estimation}, in:
  \bibinfo{booktitle}{Proc. IEEE Int. Conf. Geosci. Remote Sens. (IGARSS)}, pp.
  \bibinfo{pages}{3225--3228}.
\bibitem[{Nascimento and Dias(2005)}]{VCA}
\bibinfo{author}{Nascimento, J.}, \bibinfo{author}{Dias, J.},
  \bibinfo{year}{2005}.
\newblock \bibinfo{title}{Vertex component analysis: a fast algorithm to unmix
  hyperspectral data}.
\newblock \bibinfo{journal}{IEEE Trans. Geosci. Remote Sens.}
  \bibinfo{volume}{43}, \bibinfo{pages}{898--910}.
\bibitem[{Nielsen(2007)}]{nielsen2007regularized}
\bibinfo{author}{Nielsen, A.A.}, \bibinfo{year}{2007}.
\newblock \bibinfo{title}{The regularized iteratively reweighted {MAD} method
  for change detection in multi-and hyperspectral data}.
\newblock \bibinfo{journal}{IEEE Trans. Image Process.} \bibinfo{volume}{16},
  \bibinfo{pages}{463--478}.
\bibitem[{Nielsen et~al.(1998)Nielsen, Conradsen and
  Simpson}]{nielsen1998multivariate}
\bibinfo{author}{Nielsen, A.A.}, \bibinfo{author}{Conradsen, K.},
  \bibinfo{author}{Simpson, J.J.}, \bibinfo{year}{1998}.
\newblock \bibinfo{title}{Multivariate alteration detection ({MAD}) and {MAF}
  postprocessing in multispectral, bitemporal image data: New approaches to
  change detection studies}.
\newblock \bibinfo{journal}{Remote Sens. Environment} \bibinfo{volume}{64},
  \bibinfo{pages}{1--19}.
\bibitem[{Niu et~al.(2018)Niu, Gong, Zhan and Yang}]{niu2018conditional}
\bibinfo{author}{Niu, X.}, \bibinfo{author}{Gong, M.}, \bibinfo{author}{Zhan,
  T.}, \bibinfo{author}{Yang, Y.}, \bibinfo{year}{2018}.
\newblock \bibinfo{title}{A conditional adversarial network for change
  detection in heterogeneous images}.
\newblock \bibinfo{journal}{IEEE Geosci. Remote Sens. Lett.}
  \bibinfo{volume}{16}, \bibinfo{pages}{45--49}.
\bibitem[{Pepe(2000)}]{pepe2000receiver}
\bibinfo{author}{Pepe, M.S.}, \bibinfo{year}{2000}.
\newblock \bibinfo{title}{Receiver operating characteristic methodology}.
\newblock \bibinfo{journal}{J. Amer. Stat. Assoc.} \bibinfo{volume}{95},
  \bibinfo{pages}{308--311}.
\bibitem[{Ronneberger et~al.(2015)Ronneberger, Fischer and
  Brox}]{ronneberger2015u}
\bibinfo{author}{Ronneberger, O.}, \bibinfo{author}{Fischer, P.},
  \bibinfo{author}{Brox, T.}, \bibinfo{year}{2015}.
\newblock \bibinfo{title}{U-net: Convolutional networks for biomedical image
  segmentation}, in: \bibinfo{booktitle}{Int. Conf. Medical image Computing and
  Computer-Assisted Intervention (MICCAI)}, \bibinfo{organization}{Springer}.
  pp. \bibinfo{pages}{234--241}.
\bibitem[{Senthilkumaran and Vaithegi(2016)}]{ostu}
\bibinfo{author}{Senthilkumaran, N.}, \bibinfo{author}{Vaithegi, S.},
  \bibinfo{year}{2016}.
\newblock \bibinfo{title}{Image segmentation by using thresholding techniques
  for medical images}.
\newblock \bibinfo{journal}{Computer Science \& Engineering: An Int. Journal}
  \bibinfo{volume}{6}, \bibinfo{pages}{1--13}.
\bibitem[{Shafique et~al.(2022)Shafique, Cao, Khan, Asad and
  Aslam}]{shafique2022deep}
\bibinfo{author}{Shafique, A.}, \bibinfo{author}{Cao, G.},
  \bibinfo{author}{Khan, Z.}, \bibinfo{author}{Asad, M.},
  \bibinfo{author}{Aslam, M.}, \bibinfo{year}{2022}.
\newblock \bibinfo{title}{Deep learning-based change detection in remote
  sensing images: A review}.
\newblock \bibinfo{journal}{Remote Sens.} \bibinfo{volume}{14},
  \bibinfo{pages}{871}.
\bibitem[{Sun et~al.(2021a)Sun, Lei, Guan and Kuang}]{sun2021iterative}
\bibinfo{author}{Sun, Y.}, \bibinfo{author}{Lei, L.}, \bibinfo{author}{Guan,
  D.}, \bibinfo{author}{Kuang, G.}, \bibinfo{year}{2021}a.
\newblock \bibinfo{title}{Iterative robust graph for unsupervised change
  detection of heterogeneous remote sensing images}.
\newblock \bibinfo{journal}{IEEE Trans. Image Process.} \bibinfo{volume}{30},
  \bibinfo{pages}{6277--6291}.
\bibitem[{Sun et~al.(2020)Sun, Lei, Li, Tan and Kuang}]{sun2020patch}
\bibinfo{author}{Sun, Y.}, \bibinfo{author}{Lei, L.}, \bibinfo{author}{Li, X.},
  \bibinfo{author}{Tan, X.}, \bibinfo{author}{Kuang, G.}, \bibinfo{year}{2020}.
\newblock \bibinfo{title}{Patch similarity graph matrix-based unsupervised
  remote sensing change detection with homogeneous and heterogeneous sensors}.
\newblock \bibinfo{journal}{IEEE Trans. Geosci. Remote Sens.}
  \bibinfo{volume}{59}, \bibinfo{pages}{4841--4861}.
\bibitem[{Sun et~al.(2021b)Sun, Lei, Li, Tan and Kuang}]{sun2021structure}
\bibinfo{author}{Sun, Y.}, \bibinfo{author}{Lei, L.}, \bibinfo{author}{Li, X.},
  \bibinfo{author}{Tan, X.}, \bibinfo{author}{Kuang, G.},
  \bibinfo{year}{2021}b.
\newblock \bibinfo{title}{Structure consistency-based graph for unsupervised
  change detection with homogeneous and heterogeneous remote sensing images}.
\newblock \bibinfo{journal}{IEEE Trans. Geosci. Remote Sens.}
  \bibinfo{volume}{60}, \bibinfo{pages}{1--21}.
\bibitem[{Tewkesbury et~al.(2015)Tewkesbury, Comber, Tate, Lamb and
  Fisher}]{tewkesbury2015critical}
\bibinfo{author}{Tewkesbury, A.P.}, \bibinfo{author}{Comber, A.J.},
  \bibinfo{author}{Tate, N.J.}, \bibinfo{author}{Lamb, A.},
  \bibinfo{author}{Fisher, P.F.}, \bibinfo{year}{2015}.
\newblock \bibinfo{title}{A critical synthesis of remotely sensed optical image
  change detection techniques}.
\newblock \bibinfo{journal}{Remote Sens. Environment} \bibinfo{volume}{160},
  \bibinfo{pages}{1--14}.
\bibitem[{Vivone et~al.(2014)Vivone, Alparone, Chanussot, Dalla~Mura, Garzelli,
  Licciardi, Restaino and Wald}]{vivone2014critical}
\bibinfo{author}{Vivone, G.}, \bibinfo{author}{Alparone, L.},
  \bibinfo{author}{Chanussot, J.}, \bibinfo{author}{Dalla~Mura, M.},
  \bibinfo{author}{Garzelli, A.}, \bibinfo{author}{Licciardi, G.A.},
  \bibinfo{author}{Restaino, R.}, \bibinfo{author}{Wald, L.},
  \bibinfo{year}{2014}.
\newblock \bibinfo{title}{A critical comparison among pansharpening
  algorithms}.
\newblock \bibinfo{journal}{IEEE Trans. Geosci. Remote Sens.}
  \bibinfo{volume}{53}, \bibinfo{pages}{2565--2586}.
\bibitem[{Wald et~al.(1997)Wald, Ranchin and Mangolini}]{wald1997fusion}
\bibinfo{author}{Wald, L.}, \bibinfo{author}{Ranchin, T.},
  \bibinfo{author}{Mangolini, M.}, \bibinfo{year}{1997}.
\newblock \bibinfo{title}{Fusion of satellite images of different spatial
  resolutions: Assessing the quality of resulting images}.
\newblock \bibinfo{journal}{Photogrammetric Engineering Remote Sens.}
  \bibinfo{volume}{63}, \bibinfo{pages}{691--699}.
\bibitem[{Wang and Xu(2010)}]{wang2010comparison}
\bibinfo{author}{Wang, F.}, \bibinfo{author}{Xu, Y.J.}, \bibinfo{year}{2010}.
\newblock \bibinfo{title}{Comparison of remote sensing change detection
  techniques for assessing hurricane damage to forests}.
\newblock \bibinfo{journal}{Environmental monitoring and assessment}
  \bibinfo{volume}{162}, \bibinfo{pages}{311--326}.
\bibitem[{Wang et~al.(2021)Wang, Fu, Zeng, Sun, Zhan, Huang and
  Ding}]{wang2021enhanced}
\bibinfo{author}{Wang, W.}, \bibinfo{author}{Fu, X.}, \bibinfo{author}{Zeng,
  W.}, \bibinfo{author}{Sun, L.}, \bibinfo{author}{Zhan, R.},
  \bibinfo{author}{Huang, Y.}, \bibinfo{author}{Ding, X.},
  \bibinfo{year}{2021}.
\newblock \bibinfo{title}{Enhanced deep blind hyperspectral image fusion}.
\newblock \bibinfo{journal}{IEEE Trans. Neural Netw. Learn. Syst.} .
\bibitem[{Wang et~al.(2019)Wang, Zeng, Huang, Ding and Paisley}]{wang2019deep}
\bibinfo{author}{Wang, W.}, \bibinfo{author}{Zeng, W.}, \bibinfo{author}{Huang,
  Y.}, \bibinfo{author}{Ding, X.}, \bibinfo{author}{Paisley, J.},
  \bibinfo{year}{2019}.
\newblock \bibinfo{title}{Deep blind hyperspectral image fusion}, in:
  \bibinfo{booktitle}{Proc. IEEE Int. Conf. Computer Vision (ICCV)}, pp.
  \bibinfo{pages}{4150--4159}.
\bibitem[{Wei et~al.(2015a)Wei, Dobigeon and Tourneret}]{Wei_IEEE_JSTSP_2015}
\bibinfo{author}{Wei, Q.}, \bibinfo{author}{Dobigeon, N.},
  \bibinfo{author}{Tourneret, J.Y.}, \bibinfo{year}{2015}a.
\newblock \bibinfo{title}{Bayesian fusion of multi-band images}.
\newblock \bibinfo{journal}{IEEE J. Sel. Topics Signal Process.}
  \bibinfo{volume}{9}, \bibinfo{pages}{1117--1127}.
\bibitem[{Wei et~al.(2015b)Wei, Dobigeon and Tourneret}]{wei2015fast}
\bibinfo{author}{Wei, Q.}, \bibinfo{author}{Dobigeon, N.},
  \bibinfo{author}{Tourneret, J.Y.}, \bibinfo{year}{2015}b.
\newblock \bibinfo{title}{Fast fusion of multi-band images based on solving a
  {S}ylvester equation}.
\newblock \bibinfo{journal}{IEEE Trans. Image Process.} \bibinfo{volume}{24},
  \bibinfo{pages}{4109--4121}.
\bibitem[{Xu et~al.(2023)Xu, Shi and Zhu}]{xu2023ucdformer}
\bibinfo{author}{Xu, Q.}, \bibinfo{author}{Shi, Y.}, \bibinfo{author}{Zhu,
  X.X.}, \bibinfo{year}{2023}.
\newblock \bibinfo{title}{{UCDFormer}: Unsupervised change detection using
  real-time transformers}, in: \bibinfo{booktitle}{Proc. Joint Urban Remote
  Sensing Event (JURSE)}, \bibinfo{organization}{IEEE}. pp.
  \bibinfo{pages}{1--4}.
\bibitem[{Xu et~al.(2020)Xu, Wu, Chanussot and Wei}]{xu2020hyperspectral}
\bibinfo{author}{Xu, Y.}, \bibinfo{author}{Wu, Z.}, \bibinfo{author}{Chanussot,
  J.}, \bibinfo{author}{Wei, Z.}, \bibinfo{year}{2020}.
\newblock \bibinfo{title}{Hyperspectral images super-resolution via learning
  high-order coupled tensor ring representation}.
\newblock \bibinfo{journal}{IEEE Trans. Neural Netw. Learn. Syst.}
  \bibinfo{volume}{31}, \bibinfo{pages}{4747--4760}.
\bibitem[{Zhang et~al.(2021a)Zhang, Lin, Yang and Zhang}]{zhang2021escnet}
\bibinfo{author}{Zhang, H.}, \bibinfo{author}{Lin, M.}, \bibinfo{author}{Yang,
  G.}, \bibinfo{author}{Zhang, L.}, \bibinfo{year}{2021}a.
\newblock \bibinfo{title}{Escnet: An end-to-end superpixel-enhanced change
  detection network for very-high-resolution remote sensing images}.
\newblock \bibinfo{journal}{IEEE Trans. Neural Netw. Learn. Syst.} .
\bibitem[{Zhang et~al.(2020)Zhang, Nie, Wei, Li and Zhang}]{zhang2020deep}
\bibinfo{author}{Zhang, L.}, \bibinfo{author}{Nie, J.}, \bibinfo{author}{Wei,
  W.}, \bibinfo{author}{Li, Y.}, \bibinfo{author}{Zhang, Y.},
  \bibinfo{year}{2020}.
\newblock \bibinfo{title}{Deep blind hyperspectral image super-resolution}.
\newblock \bibinfo{journal}{IEEE Trans. Neural Netw. Learn. Syst.}
  \bibinfo{volume}{32}, \bibinfo{pages}{2388--2400}.
\bibitem[{Zhang et~al.(2021b)Zhang, Huang, Wang and Li}]{SSRNET}
\bibinfo{author}{Zhang, X.}, \bibinfo{author}{Huang, W.},
  \bibinfo{author}{Wang, Q.}, \bibinfo{author}{Li, X.}, \bibinfo{year}{2021}b.
\newblock \bibinfo{title}{Ssr-net: Spatial–spectral reconstruction network
  for hyperspectral and multispectral image fusion}.
\newblock \bibinfo{journal}{IEEE Trans. Geosci. Remote Sens.}
  \bibinfo{volume}{59}, \bibinfo{pages}{5953--5965}.
\bibitem[{Zhao et~al.(2016)Zhao, Wei, Basarab, Dobigeon, Kouam{\'e} and
  Tourneret}]{zhao2016fast}
\bibinfo{author}{Zhao, N.}, \bibinfo{author}{Wei, Q.},
  \bibinfo{author}{Basarab, A.}, \bibinfo{author}{Dobigeon, N.},
  \bibinfo{author}{Kouam{\'e}, D.}, \bibinfo{author}{Tourneret, J.Y.},
  \bibinfo{year}{2016}.
\newblock \bibinfo{title}{Fast single image super-resolution using a new
  analytical solution for $\ell_{2} - \ell_{2}$ problems}.
\newblock \bibinfo{journal}{IEEE Trans. Image Process.} \bibinfo{volume}{25},
  \bibinfo{pages}{3683--3697}.
\bibitem[{Zhao et~al.(2017)Zhao, Wang, Gong and Liu}]{zhao2017discriminative}
\bibinfo{author}{Zhao, W.}, \bibinfo{author}{Wang, Z.}, \bibinfo{author}{Gong,
  M.}, \bibinfo{author}{Liu, J.}, \bibinfo{year}{2017}.
\newblock \bibinfo{title}{Discriminative feature learning for unsupervised
  change detection in heterogeneous images based on a coupled neural network}.
\newblock \bibinfo{journal}{IEEE Trans. Geosci. Remote Sens.}
  \bibinfo{volume}{55}, \bibinfo{pages}{7066--7080}.

\end{thebibliography}
\end{document}